\documentclass[aps,nofootinbib,superscriptaddress, showpacs,preprintnumbers,  nofootinbibt,onecolumn]{revtex4-2}
\usepackage{epsfig}
\usepackage{multirow}
\usepackage{eurosym}
\usepackage{dcolumn}
\usepackage{bm}
\usepackage{enumerate}
\usepackage{float}
\usepackage{epstopdf}
\usepackage{amsmath}
\usepackage{bm}
\usepackage{amsfonts}
\usepackage{amssymb}
\usepackage{graphicx}
\usepackage{alphalph,mathtools}
\usepackage{etoolbox}
\usepackage{color}
\usepackage{booktabs}
\usepackage{footnote}
\usepackage{makecell,tabularx}
\usepackage{hyperref}

\hypersetup{colorlinks,citecolor=blue}
\hypersetup{colorlinks=true,linkcolor=red,filecolor=magenta,    urlcolor=blue}

\def\be{\begin{equation}}
\def\ee{\end{equation}}
\def\bea{\begin{eqnarray}}
\def\eea{\end{eqnarray}}

\begin{document}

\title{Model-independent study for a quintessence model of dark energy: Analysis and Observational constraints}
\author{Amine Bouali}
\email{a1.bouali@ump.ac.ma}
\affiliation{Laboratory of Physics of Matter and Radiation, Mohammed I University, BP 717, Oujda, Morocco}
\author{Himanshu Chaudhary}
\email{himanshuch1729@gmail.com}
\affiliation{Department of Applied Mathematics, Delhi Technological University, Delhi-110042, India} 
\affiliation{Pacif Institute of Cosmology and Selfology (PICS), Sagara, Sambalpur 768224, Odisha, India}
\affiliation{Department of Mathematics, Shyamlal College, University of Delhi, Delhi-110032, India.}
\author{Amritansh Mehrotra}
\email{amritanshmehrotra71@gmail.com}
\affiliation{Department of Power and Electrical, University of Petroleum and Energy Studies, Dehradun, Uttarakhand, India}
\author{S. K. J. Pacif}
\email{shibesh.math@gmail.com}
\affiliation{Pacif Institute of Cosmology and Selfology (PICS), Sagara, Sambalpur 768224, Odisha, India} 
\affiliation{Centre for Cosmology and Science Popularization (CCSP), SGT University, Delhi-NCR, Gurugram 122505, Haryana, India}

\begin{abstract}
In this paper, a well-motivated parametrization of the Hubble parameter ($H$%
) is revisited that renders two models of dark energy showing some
intriguing features of the late-time accelerating Universe. A general
quintessence field is considered as a source of dark energy. We have
obtained tighter constraints using recently updated cosmic observational
datasets for the considered models. The two models described here show a
nice fit to the considered uncorrelated Hubble datasets, Standard candles,
Gamma Ray Bursts, Quasars, and uncorrelated Baryonic Acoustic Oscillations
datasets. Using the constrained values of the model parameters, we have
discussed some features of the late-time accelerating models and obtained
the present value of the deceleration parameter ($q_{0}$), the present value of
the Hubble parameter ($H_{0}$) and the transition redshift ($z_{t}$) from
deceleration to acceleration. The current value of the deceleration parameter   for both models is consistent with the Planck 2018 results. The evolution of the geometrical
and physical parameters is discussed through graphical representations
for both models with some diagnostic analysis. The statistical analysis
performed here shows greater results and overall, the outcomes of this
investigation are superior to those previously found.
\end{abstract}

\maketitle
\tableofcontents


\section{Introduction}

\label{Introduction} The most debated topics of modern-day cosmology is the
the late-time accelerating expansion of the Universe. The early evidence of
accelerating expansion of the Universe was provided by the observation of
high redshift data of supernovas of type Ia (SNIa) \cite{1, 2}. Later on, by
virtue of improved cosmological measurement techniques \cite{3, 4, 5},
rigorous analysis and precise observations indicated accelerated cosmic
inflation. Indirect evidence of a nonzero cosmological constant was provided
by the data from the Cosmic Microwave Background Radiation (CMB) and Large
Scale Structure (LSS) \cite{1, 6, 7}. The current cosmological research is
more exciting with the release of ongoing high-precision data in the fields
of cosmology and astrophysics in the past few decades. One of the biggest
issues now in this field is the understanding of the Universe's accelerated
expansion phase that produced a plethora of cosmological models to
describe the observational datasets released in the past few years such as
SNIa, acoustic peaks of CMB, Observational Hubble datasets, Baryon acoustic
oscillations (BAO), strong lensing systems (SLS) \cite{5}, and a few more.
The simplest one is of course the famous $\Lambda $CDM (Cold Dark Matter)
model, which explains these cosmological observations very well. The model
includes the cosmological constant ($\Lambda $ or CC), which is
characterized by its constant equation of state (EoS) $w=-1$ to model the
accelerated expansion of the Universe and dust matter $w=0$ to simulate the
evolution of dark matter at the background level. These two extra components
correspond to approximately 95 percent of the whole energy budget in the
Universe (the rest of the components are associated with baryons and
relativistic species like photons and neutrinos).In literature, these two
ingredients are referred to as dark energy (DE) and dark matter (DM).
Detection and observation of DM and DE is another problem that is still
persistent in cosmology. Besides its success at a large scale, $\Lambda $CDM
presents several issues at local scales, for instance, the well-known
missing satellite problem that refers to the discrepancy of about 10 times
more dwarf galaxies obtained by the numerical simulations based on the $%
\Lambda $CDM model and the observed ones in a cluster of galaxies \cite{6, 7}%
. Also, the well-known core-cusp problem \cite{1}. In this sense, the
concordance issue is the discrepancy between the CC value measured from the
the perspective of quantum field theory and the one derived from cosmological
measurements, which is about 120 orders of magnitude. \cite{8, 9, 10}.
Additionally, the degeneracy problem which afflicts also the $\Lambda $CDM
model refers to the inability of measuring the energy-momentum of each
component, instead of the total one. In other words, this suggests that it
is difficult to determine whether the dark region is made up of one or more
components. Several models have emerged in order to propose alternatives to
the $\Lambda $CDM paradigm, for instance, Chaplygin gas \cite{11}, and
Unimodular gravity \cite{12, 13}, among others \cite{17}, have entered into
the scene as a greater contender, resolving conundrums that the $\Lambda $%
CDM cannot. Moreover, scalar fields such as DM \cite{15, 16, 17, 18}, axion 
\cite{19, 20}, etc, are important approaches to resolving the problem of DM.
Fluids with viscosity are great candidates not only to aboard the DM problem
but also, the DE problem from a unifying approach. However, a general scalar
field as a candidate for dark energy is a good approach that can also
describe the CC. Tillto variety of dynamically evolving scalar
field models has been thought of which include quintessence \cite{17, 18, 19}%
, K-essence \cite{20, 21, 22}, phantom \cite{23}, and tachyonic fields \cite%
{24, 25, 26}. To understand the behavior of our physical Universe, finding
the exact solutions of Einstein's Field Equations (EFEs) are necessary. A
lot of work was done by many theoreticians in finding the solution to EFEs
right after the formulation of EFEs. The Schwarchild solution being the
first solution and the perfect fluid equations were treated as an additional
condition. Various other exact solutions were obtained for static and
spherically symmetric metrics like Einstein's static solution, de-Sitter
solution \cite{27, 28}, Tolman's solution \cite{29}, Adler's solution \cite%
{30}, Buchdahl's solution \cite{31}, Vaidya and tikekar solution \cite{32},
Durgapal's solution \cite{33}, Knutsen's solution \cite{34}, and many more.
All these solutions were obtained even though EFEs are highly
nonlinear. All the models mentioned above theoretically explain
our Universe very well. Apart from the theoretical verification the
observations also play an important role. Observations either validate a
model or rejects it based on numerical computations and also
parametrization of these models. Therefore it is essential to discuss
observational datasets when making a theoretical model of the Universe.
Before 1998, scientists were trying to construct a model with $\Lambda $,
which can resolve the age crisis problem. All these models can also be found
in the reference \cite{35} Recently, several theoretical models have been developed to analyze the entire evolutionary history of the Universe through parametrization of $q(z)$ as a function of scale factor $(a(t))$ or time $(t)$ or redshift $(z)$
in the reference \cite{him1,him2,him3,him4,him5,him6,him7,him8,him9}. The following is how this document is organized. In the introduction part,
models in modern cosmology are discussed. In the next section, Einstein's field equation in general relativity taking into consideration dark matter
has been discussed. The third section deals with finding the exact solution of EFEs and also the parametrization of different models that have been
taken into account.

\section{Field Equations with Quintessence}

Dark energy (DE) has emerged as one of the most crucial mysteries throughout
the domain of cosmology, and it is a subject of debate whether it can be
characterized as a source term in Einstein's field equations. Quintessence
is one of the notable candidates of dark energy after the cosmological
constant, that is compatible with various cosmological observations. It's
more specifically a scalar field, that is proposed in physics to account for
the Universe's observed accelerated pace of expansion. The initial
illustration of this situation was put out in \cite{36}. This notion was
then expanded to more general varieties of time-varying dark energy models
(e.g. K-essence, Spint-essence etc.) for which R. R. Caldwell et al. \cite%
{37} had to introduce the term \textquotedblleft species of
essence\textquotedblright . Quintessence may be attractive or repulsive
depending on the ratio of kinetic and potential energies of the field. The
action that represents our physical system is,

\begin{equation}
S=\int d^{4}x\sqrt{-g}\left( \frac{1}{2}M_{Pl}^{2}R+L_{m}+L_{\phi }\right) 
\text{,}  \label{Act1}
\end{equation}%
where $L_{m}$ is the Lagrangian of matter fields and $L_{\phi }$ is the
Lagrangian of the scalar field given by,%
\begin{equation}
L_{\phi }=-\frac{1}{2}g^{\mu \nu }\partial _{\mu }\phi \partial _{\nu }\phi
-V\left( \phi \right) \text{,}  \label{Act2}
\end{equation}%
where $g$ is the determinant of the metric $g_{\mu \nu }$, $M_{Pl}=\left(
8\pi G\right) ^{-1/2}$ is the reduced Planck mass, $R$ is the Ricci scalar,
and $V\left( \phi \right) $ is a general self-coupling potential for which $%
\phi $ must be positive for physically acceptable fields. We assume that
non-relativistic matter does not have a direct coupling to the quintessence
field (minimal interaction). The differentiation with respect to $g_{\mu \nu
}$ leads to the gravitational field equations,

\begin{equation}
R_{\mu \nu} - \frac{1}{2}Rg_{\mu \nu} = M_{Pl}^{-2} T_{\mu \nu}^{\text{Total}} \text{.} \label{Act3}
\end{equation}

where 
\begin{equation}
T_{\mu \nu }^{\phi }=\partial _{\mu }\phi \partial _{\nu }\phi -\frac{1}{2}%
g_{\mu \nu }\left( \partial \phi \right) ^{2}-g_{\mu \nu }V\left( \phi
\right) \text{,}  \label{Act4}
\end{equation}%
is the energy-momentum tensor of the scalar field. Here, $\left( \partial \phi
\right) ^{2}\equiv g_{\alpha \beta }\partial ^{\alpha }\phi \partial ^{\beta
}\phi $ and $T_{\mu \nu }^{Total}=T_{\mu \nu }^{M}+T_{\mu \nu }^{\phi }$.
The differentiation with respect to $\phi $ leads to the Klein-Gordon
equation $\triangledown _{\mu }\triangledown ^{\mu }\phi -\frac{\partial V}{%
\partial \phi }=0$.

We start by considering the most general homogeneous and isotropic
space-time, which is the Friedmann-Lema\^{\i}tre-Robertson-Walker (FLRW)
metric, 
\begin{equation}
ds^{2}=-dt^{2}+a(t)^{2}\left( \frac{dr^{2}}{1-kr^{2}}+r^{2}d\theta
^{2}+r^{2}\sin ^{2}\theta d\phi ^{2}\right) ,  \label{3}
\end{equation}%
where $k=1,$ $0,$ $-1$ for closed, flat and open geometries respectively. We
have taken the velocity of light $c=1$. Analysis suggests the topology of the
the Universe is flat and we have the space-time,

\begin{equation}
ds^{2}=-dt^{2}+a^{2}(t)\left[ dr^{2}+r^{2}\left( d\theta ^{2}+\sin
^{2}\theta d\phi ^{2}\right) \right] ,  \label{4}
\end{equation}%
here $a(t)$ is the Universe's scale factor. In this background, the Einstein
field equations yield the following dynamical equations, 
\begin{equation}
M_{Pl}^{-2}\rho _{Total}=3\left( \frac{\dot{a}}{a}\right) ^{2}=3H^{2},
\label{eq5}
\end{equation}%
\begin{equation}
M_{Pl}^{-2}p_{Total}=-2\frac{\ddot{a}}{a}-\left( \frac{\dot{a}}{a}\right)
^{2}=(2q-1)H^{2}.  \label{eq6}
\end{equation}%
Here $H\left( =\frac{\dot{a}}{a}\right) $ is the Hubble parameter and $%
q\left( =-\frac{a\ddot{a}}{\dot{a}^{2}}\right) $ is the deceleration
parameter, which are higher order derivatives of the scale factor $a$ and
determine the dynamics of the Universe. The total energy density $\rho
_{Total}=\rho _{M}+\rho _{\phi }$ and $p_{Total}=p_{M}+p_{\phi }$. The
energy density and pressure of the scalar field are: $\rho _{\phi }=\frac{1}{%
2}\dot{\phi}^{2}+V(\phi )$, $p_{\phi }=\frac{1}{2}\dot{\phi}^{2}-V(\phi )$.
Therefore, the equation of state (EoS) of the scalar field (dark energy)
will be%
\begin{equation}
\omega _{\phi }=\frac{p_{\phi }}{\rho _{\phi }}=\frac{\frac{1}{2}\dot{\phi}%
^{2}-V(\phi )}{\frac{1}{2}\dot{\phi}^{2}+V(\phi )}\text{.}  \label{Eosqf}
\end{equation}

We must remember that the EoS of dark energy (quintessence) is a dynamically
evolving parameter that can take values in the range $\left[ -1,1\right] \,$%
. We can see for a potential dominated field, $\omega _{\phi }$ reduce to $%
-1 $ i.e. cosmological constant recovered. For $\omega _{\phi }$ crossing
the value $-1$ describes the phantom, which is beyond our scope here.

One could yield the conservation equation from (\ref{eq5}) and (\ref{eq6}), 
\begin{equation}
\dot{\rho}_{Total}+3\left( p_{Total}+\rho _{Total}\right) \frac{\dot{a}}{a}%
=0.  \label{eq7}
\end{equation}

The conservation equation (\ref{eq7}) is important in evolution since it
deals with matter and its interactions. In modern cosmology, two types of
dark energy models are frequently discussed: interacting dark energy models
(covering the connection between cold dark matter and dark energy) \cite{38,
39, 40}, and non-interacting dark energy models (allowing all matters to
evolve separately) \cite{35, 41, 42, 43}. There is currently no recognized
interaction between matter and DE other than gravity. The current study
exclusively considers non-interacting models. The equations in the system
are nonlinear ordinary differential equations, and explicit solutions are
challenging to find. In the past, enormous attempts were made to develop
both precise and numeric solutions to EFEs. The solution strategies for the
above-mentioned system of equations will be explained in detail in the
following section.

\section{Solution of Field Equations in a Model-independent way}

In the above derived equations, there are only two independent equations with four unknowns $a,\rho $, $p$, or $\omega $ in the above system of equations (\ref{eq5}), (\ref{eq6}), and (%
\ref{eq7}). Because of the homogenous matter distribution on a wide scale throughout the Universe, it is conventional to analyze the barotropic equation of state, $p=\omega \rho $, $\omega \in \lbrack 0,1]$. Depending on the discrete or dynamical values of the EoS parameter. The EoS defines various kinds of matter sources in the Universe e.g. for $\omega =0$, dust-like matter; $\omega =1/3$, radiation; $\omega =-1$, etc. The third constraint equation would be provided by this extra equation. However, the linear equation of state is not the only choice and there can be a complicated equation of state depending upon the matter source. with the inclusion of the extra source of energy in the Universe in order to explain the current accelerating expansion of the Universe, the field equations are burdened with one extra variable resulting in the deficiency of one more equation to close the system for a consistent solution. Therefore, the other constraint equation (the fourth one) would be the consideration of the EoS
of DE $\omega _{de}=p_{de}/\rho _{de}$, function of time $t$, scale factor $a$ or redshift $z$), also referred as \textit{parametrization of dark energy equation of state}. In literature, there are various forms of parametrization of dark energy equation of state e.g. Chaplygin gas equation of state, Modified Chaplygin, variable Chalplygin gas equation of state,  polytropic gas equation of state, logotropic gas equation of state, Vander Waal's equation of state, etc. With these four equations, one could describe the cosmological dynamics of the Universe by presenting all other geometrical parameters such as Hubble parameter ($H$), deceleration parameter ($q$), jerk parameter ($j$), etc., or physical parameters such as energy density of matter or dark energy ($\rho $), dark energy pressure ($p$), equation of state of matter or dark energy ($\omega $), etc. as functions of cosmic time ($t$), scale factor ($a$) or redshift ($z$).\\\\ 
Besides the parametrization of the dark energy equation of state, there are also possibilities of considering any functional forms of the scale factor $a$. There are numerous strategies in the literature for parametrizing the scale factor and its higher order derivatives $H$, $q$, $j$ too, which provide the entire solution of the EFEs, i.e. the explicit forms of cosmological parameters as a function of cosmic time $t$. In fact, there have been two different approaches to analyzing the solution of EFEs in general relativity theory or on some modified theories: one is through
parametrization of the \textit{geometrical} parameters $a$, $H$, $q$ or $j$ as a function of cosmic time; the second is through parametrization of \textit{physical} parameters $\rho $, $p$, or $\omega $) as a function of scale factor or redshift. If we examine closely, we might remark that the primary type of parametrization of geometrical parameters is studied to produce exact
solutions that address the expanding dynamics of the Universe and give the
time evolution of the physical parameters $\rho $, $p$, or $\omega $. This
approach is also known as the model-independent method of studying
cosmological models or cosmological parametrization \cite{44}, \cite{Pacif-parametrization-1}, \cite{Pacif-parametrization-2}. The model-independent way has  the potential of rebuilding the cosmic history of the Universe as well as interpreting some of the Universe's phenomena. The beauty of this approach is that this does not affect the background theory and provides the simplest mathematical way to reconstruct the cosmic history of the Universe. Furthermore, this strategy gives the easiest way to theoretically overcome a few problems of the Standard model, including the initial singularity problem, cosmological constant problem, and the all-time decelerated expansion issue of the Standard model. Whereas the second type of parametrization of
physical parameters is commonly used to explain various physical features of the
Universe.
By employing any scheme of cosmological parametrization, one can encapsulate these field equations into a manageable set of parameters, allowing for a more streamlined and analytically tractable description. These free parameters (model parameters) can be constrained through any observational data. So, this approach provides a valuable tool for testing different theoretical ideas, refining models, and constraining the values of parameters through observational constraints. There are a few intriguing models of dark energy and modified gravity based on various parametrization schemes of some geometrical parameters \cite{Pacif-parametrization-3}, \cite{Pacif-parametrization-4}, \cite{Pacif-parametrization-5}, \cite{Pacif-parametrization-6}, \cite{Pacif-parametrization-7}. This discussion motivates us to consider a specific form of Hubble parameter. The goal of this paper is to find an exact solution to the EFE in conventional GR theory using a simple parametrization of the Hubble parameter $H(t)$ and reconstruct the cosmic evolution.\\\\
The cosmographic analysis gives insights into analyzing the evolution of the observable Universe in terms of kinematic factors inside a model-independent way \cite{46} idea. Furthermore, the investigation of cosmographic parameters assists mostly in the study of dark energy without the necessity of any specific cosmological model beyond the cosmological principle. The scale factor could be extended in Taylor's series around the present time $t_{0}$ in the
conventional approximation, which corresponds to the straightforward
the technique used in the cosmographic analysis. The Taylor's series expansion
could be expressed as:

\begin{equation*}
a^{(n)}=  1+H_{0}\left(t-t_{0}\right)-\frac{1}{2 !} q_{0}
H_{0}^{2}\left(t-t_{0}\right)^{2}+\frac{1}{3 !} j_{0}
H_{0}^{3}\left(t-t_{0}\right)^{3} +\frac{1}{4 !} s_{0}
H_{0}^{4}\left(t-t_{0}\right)^{4}+\frac{1}{5 !} l_{0}
H_{0}^{5}\left(t-t_{0}\right)^{5}+\ldots . . 
\end{equation*}

Here $H(t)=\frac{1}{a}\frac{da}{dt}$ is Hubble parameter, $q(t)=-\frac{1}{a}%
\frac{d^{2}a}{dt^{2}}\left[ \frac{1}{a}\frac{da}{dt}\right] ^{-2}$ is
deceleration parameter, $j(t)=\frac{1}{a}\frac{d^{3}a}{dt^{3}}\left[ \frac{1%
}{a}\frac{da}{dt}\right] ^{-3}$ is jerk parameter, $s(t)=\frac{1}{a}\frac{%
d^{4}a}{dt^{4}}\left[ \frac{1}{a}\frac{da}{dt}\right] ^{-4}$ is snap
parameter and $l(t)=\frac{1}{a}\frac{d^{5}a}{dt^{5}}\left[ \frac{1}{a}\frac{%
da}{dt}\right] ^{-5}$ is lerk parameter. All of these parameters play
significant roles in cosmographic analysis. Motivated by the preceding
arguments, this study considers a basic parametrization of the Hubble
parameter, $H(z)$, as an explicit function of cosmic time `$t$' in the form 
\cite{45},

\begin{equation}
H(t)=\frac{a_{2}t^{n}}{\left( t^{m}+a_{1}\right) ^{b}}\text{,}  \label{8}
\end{equation}%
where $a_{1},a_{2}\neq 0,n,m,b$ are real constants. $a_{1}$ and $a_{2}$ both
have time dimensions. Several specific values of the parameters $n,m,b$
suggest several interesting models, which are elaborated by Pacif et al. 
\cite{45}. Out of the twelve models for some integer values of these
constants, two models show transitions from early deceleration. In what follows, we study the aforementioned   models that we label Model 1 and Model 2 which seem to be suitable  to describe the current cosmic acceleration. 

\subsection{Model 1}
As a first, model we consider the following form:
\begin{equation}
H(t)=\frac{a_{2}}{t\left( a_{1}-t\right) , }  \label{9}
\end{equation}%
which yield%
\begin{equation}
a(t)=\zeta \left( \frac{t}{a_{1}-t}\right) ^{\frac{a_{2}}{a_{1}}}\text{, }%
\zeta \text{ is an integrating constant.}  \label{10}
\end{equation}%
It is convenient to introduce the cosmological parameters as functions of
redshift $z$. The time-redshift correlation must be constructed because the
cosmological parameters here are functions of cosmic time t. The $t-z$
relations are determined as follows: 
\begin{equation}
t(z)=a_{1}\left[ 1+\{\zeta (1+z)\}^{\frac{a_{1}}{a_{2}}}\right] ^{-1},
\label{eq11}
\end{equation}
For Model 1. The expression (\ref{eq11}) contains three parameters $\zeta $, 
$a_{1}$ and $a_{2}$, but it is sufficient to describe these models
parameters by taking $\frac{a_{1}}{a_{2}}=\gamma $, which is also beneficial
for further analysis and numerical computations for which the expressions
for the Hubble parameter could be written in terms of redshift $z$,

\begin{equation}
H(z)=H_{0}\left( 1+\zeta ^{\gamma }\right) ^{-2}(1+z)^{-\gamma }\left[
1+\{\zeta (1+z)\}^{\gamma }\right] ^{2}\text{.}  \label{12}
\end{equation}

\subsection{Model 2}

As a second  model, we consider the following form:

\begin{equation}
H(t)=\frac{a_{2}}{t\left( a_{1}-t^{2}\right) }\text{ and }a(t)=\zeta \left( 
\frac{t^{2}}{a_{1}-t^{2}}\right) ^{\frac{a_{2}}{2a_{1}}}\text{.}  \label{13}
\end{equation}

The $t-z$ relations are determined as follows:

\begin{equation}
t(z)=\sqrt{a_{1}}\left[ 1+\{\zeta (1+z)\}^{2\frac{a_{1}}{a_{2}}}\right] ^{-%
\frac{1}{2}}\text{.}  \label{15}
\end{equation}

The Hubble parameter in terms of redshift $z$ is,

\begin{equation}
H(z)=H_{0}\left( 1+\zeta ^{2\gamma }\right) ^{-\frac{3}{2}}(1+z)^{-2\gamma }%
\left[ 1+\{\zeta (1+z)\}^{2\gamma }\right] ^{\frac{3}{2}}\text{.}  \label{16}
\end{equation}

Now, we must confront these models with some cosmological data. In the following section, we consider some external datasets and find the best fit
values of the model parameters for further analysis.

\section{Data description with Results}


\label{DD} Throughout this part, we will be using three distinct types of
observational datasets to limit our model parameters together with the CMB observations. We utilized the $H(z)$
datasets of $57$ measurements, the Pantheon dataset of $1048$, $17$ uncorrelated BAO measurements, and CMB measurements in order to achieve the optimal value for the
proposed model parameters. To construct the Markov Chain Monte Carlo (MCMC) 
\cite{47}, we used the open-source tools Polychord \cite{48} and GetDist 
\cite{49}. The total $\chi ^{2}$ function of the combination $H(z)$ +
Pantheon + BAO + CMB and define as,

\begin{equation}
\chi _{tot}^{2}=\chi _{H(z)}^{2}+{\chi }_{SN}^{2}+{\chi }_{BAO}^{2}+{\chi }%
_{CMB}^{2}\text{.}
\end{equation}

\subsection{$H(z)$ Dataset}

Various observational constraints were used in order to get significant
constraints on the model parameters. We employ the $H(z)$ measurements in
our investigation to constrain the model parameters. Hubble data may be
determined in general by estimating the BAO in the radial direction of
galaxy clustering \cite{50}, or by the differential age technique, which
also provides the redshift dependency of the Hubble parameter as

\begin{equation}
H(z)=-\frac{1}{1+z}\frac{dz}{dt},
\end{equation}%
where $dz$/$dt$ is computed using two moving galaxies in a proportionate
manner. $57$ Hubble measurements were taken into the study to estimate the
parameters for the model, which are spans throughout a redshift range of $%
0.07\leqslant z\leqslant 2.42$. For comparing the observational and
theoretical observations Chi-Square function is used.

\begin{equation}
\chi_{H}^{2} \\
=\sum_{i=1}^{57} \frac{\left[H_{t h}\left(z_{i},\right)-H_{o b
s}\left(z_{i}\right)\right]^{2}}{\sigma_{H\left(z_{i}\right)}^{2}},
\end{equation}

where $H_{th\text{ }}$, $H_{obs}$ and $\sigma _{H\left( z_{i}\right) }$
denotes the model prediction, observed value of Hubble rate, and standard
error at the redshift $z_{i}$, respectively. The Hubble function numerical
values for the appropriate redshifts are shown in (see table I in \cite{51}).

\subsection{Standard Candles}

The measurement of type Ia supernovae (SNIa) determines the cosmic
accelerated expansion. Until now, SNIa has proven to be one of the most
substantial and successful methods for studying the nature of dark energy.
In recent years, several supernova data sets have been established \cite{52,
53, 54, 55, 56}. The Pantheon sample has recently been updated.\cite{57}.
The former dataset contains $1048$ spectroscopically verified SNIa spanning
in the redshift range of $0<z<2.3$. SNIa are also astronomical objects that
act as standard candles for determining relative distances. As a
consequence, SN Ia samples are combined with the distance modulus $\mu =m-M$%
, where $m$ indicates a certain object's apparent magnitude of a specific
SNIa. The chi-square of the SNIa measurements is given by,

\begin{equation}
{\chi}_{SN}^2={\Delta \mu}^{T}\hspace{0.1cm}.\hspace{0.1cm}\mathbf{C}%
_{SN}^{-1}\hspace{0.1cm}.\hspace{0.1cm}{\Delta \mu} .
\end{equation}

$\mathbf{C}_{SN}$ is represented by a covariance matrix, and ${\Delta \mu }%
=\mu _{obs}-\mu _{th}$, where $\mu _{obs}$ signifies the measured distance
modulus of a certain SNIa, meanwhile the theoretical distance modulus is
represented as $\mu _{th}$, and calculated as,

\begin{equation}
\mu _{th}(z)=5\log _{10}\frac{D_{L}(z)}{(H_{0}/c)Mpc}+25\text{.}
\end{equation}

Here $H_{0}$ signifies the current Hubble rate and $c$ reflects the speed of
light. For the flat Friedmann-Lemaitre-Robertson-Walker (FLRW) Universe, the
luminosity distance, $D_{L}$, is outlined as follows :

\begin{equation}
D_{L}(z)=(1+z)H_{0}\int_{0}^{z}\frac{dz^{\prime }}{H\left( z^{\prime
}\right) }\text{.}
\end{equation}

Because we limit the model's free parameters at the same time, i.e. by using
the Pantheon sample, hence

\begin{equation}
{\chi }_{SN}^{2}={\Delta \mu }^{T}\times \mathbf{C}_{Pantheon}^{-1}\times {%
\Delta \mu }\text{.}
\end{equation}

Throughout this, the Quasars \cite{58} and Gamma Ray Bursts \cite{59} have
also been taken into consideration.

\subsection{Baryon Acoustic Oscillations}

From the largest dataset of 333 BAO measurements, we selected 17
uncorrelated BAO measurements because using the full BAO catalog could
result in a very significant error due to data correlations. Therefore, in
order to reduce errors, we chose a limited dataset. (see table I in \cite{60}
) from \cite{61, 62, 63, 64, 65, 66, 67, 68, 69, 70, 71, 72}. Studies of the
BAO featured in the transverse direction provides a measurement of $%
D_{H}(z)/r_{d}=c/H(z)r_{d}$, where $r_{d}$ is the sound horizon at the drag
epoch and it is taken as an independent parameter and with the co-moving
angular diameter distance \cite{73, 74} being

\begin{equation}
D_M=\int_0^z \frac{c d z^{\prime}}{H\left(z^{\prime}\right)} .
\end{equation}

In our database, we also use the angular diameter distance $D_A=D_M /(1+z)$
and $D_V(z) / r_d$, which is a combination of the BAO peak coordinates
above, namely

\begin{equation}
D_V(z) \equiv\left[z D_H(z) D_M^2(z)\right]^{1 / 3} .
\end{equation}

\subsection{Cosmic Microwave Background}

\label{CMB}

The CMB distant prior measurements are taken \cite{75}. The distance priors
offer useful details about the CMB power spectrum in two ways: the acoustic
scale $l_{A}$ characterizes the CMB temperature power spectrum in the
transverse direction, causing the peak spacing to vary, and the "shift
parameter" $R$ influences the CMB temperature spectrum along the
line-of-sight direction, affecting the peak heights, which are defined as
follows:

\begin{equation}
l_A=(1+z_d) \frac{\pi D_A(z)}{r_s},
\end{equation}

\begin{equation}
\quad R(z)=\frac{\sqrt{\Omega_m} H_0}{c}(1+z_d) D_A(z)
\end{equation}

The observables that \cite{75} reports are:$R_{z}=1.7502\pm 0.0046,\quad
l_{A}=301.471\pm 0.09,\quad n_{s}=0.9649\pm 0.0043$ and $r_{s}$ is an
independent parameter, with an associated covariance matrix. (see table I in 
\cite{75}). The points represent the inflationary observables as well as the
CMB epoch expansion rate. In addition to the CMB points, we also take into
account other data from the late Universe. The result is a successful test
of the model in relation to the data.\\\\

The contour plots for the combined result of $H(z)$ + SNIa + GRB + Q + BAO + CMB are shown in the following Fig:- \ref{fig1} \& \ref{fig2} and the best-fit values with error bars are tabulated in Table \ref{tab1}.   

\begin{figure}[!htb]
   \begin{minipage}{0.49\textwidth}
     \centering
   \includegraphics[scale=0.55]{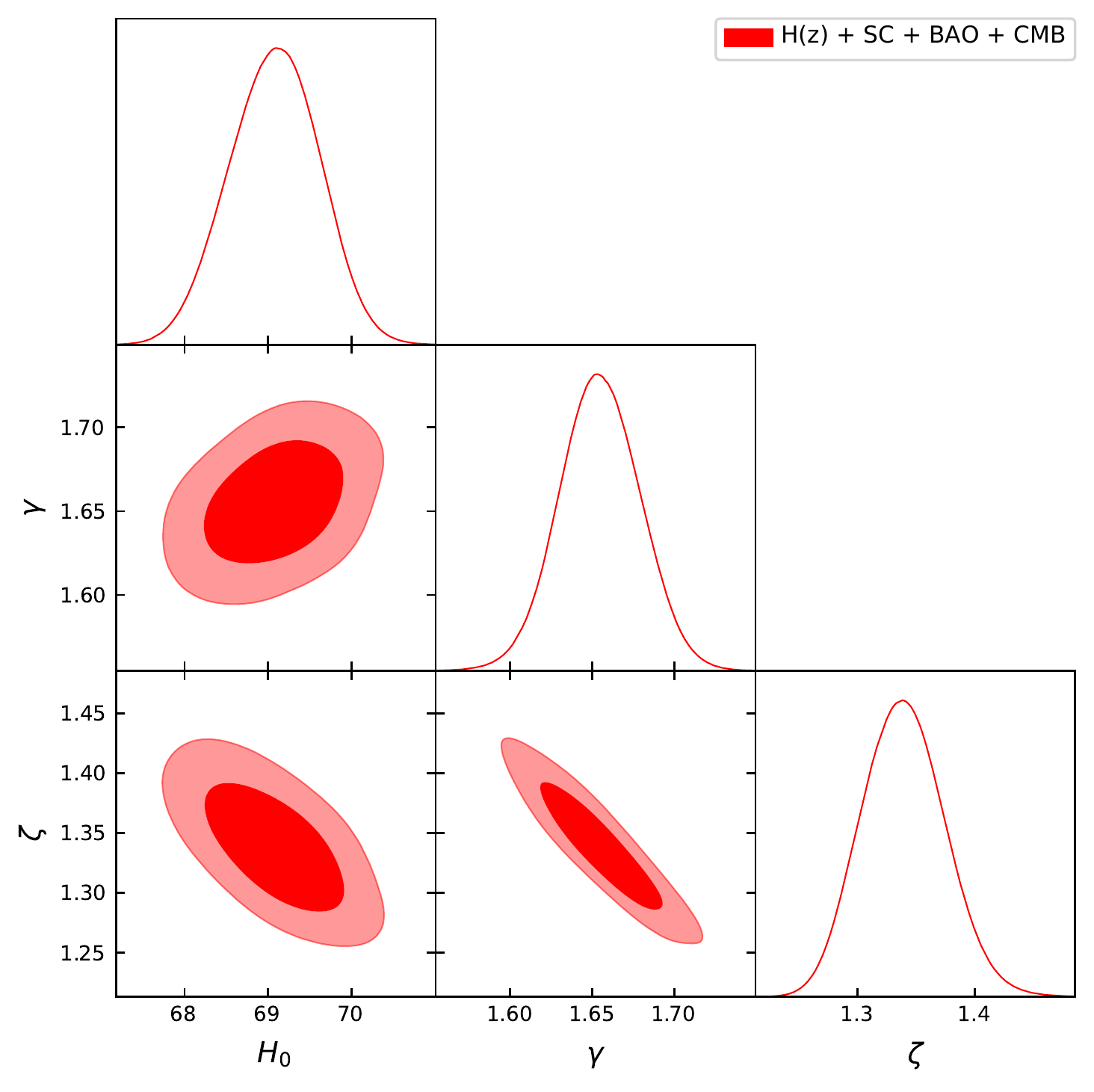}
\caption{MCMC confidence contours at 1$\protect\sigma$ and 2$\protect\sigma$ for Model 1.}\label{fig1}
   \end{minipage}\hfill
   \begin{minipage}{0.49\textwidth}
     \centering
    \includegraphics[scale=0.55]{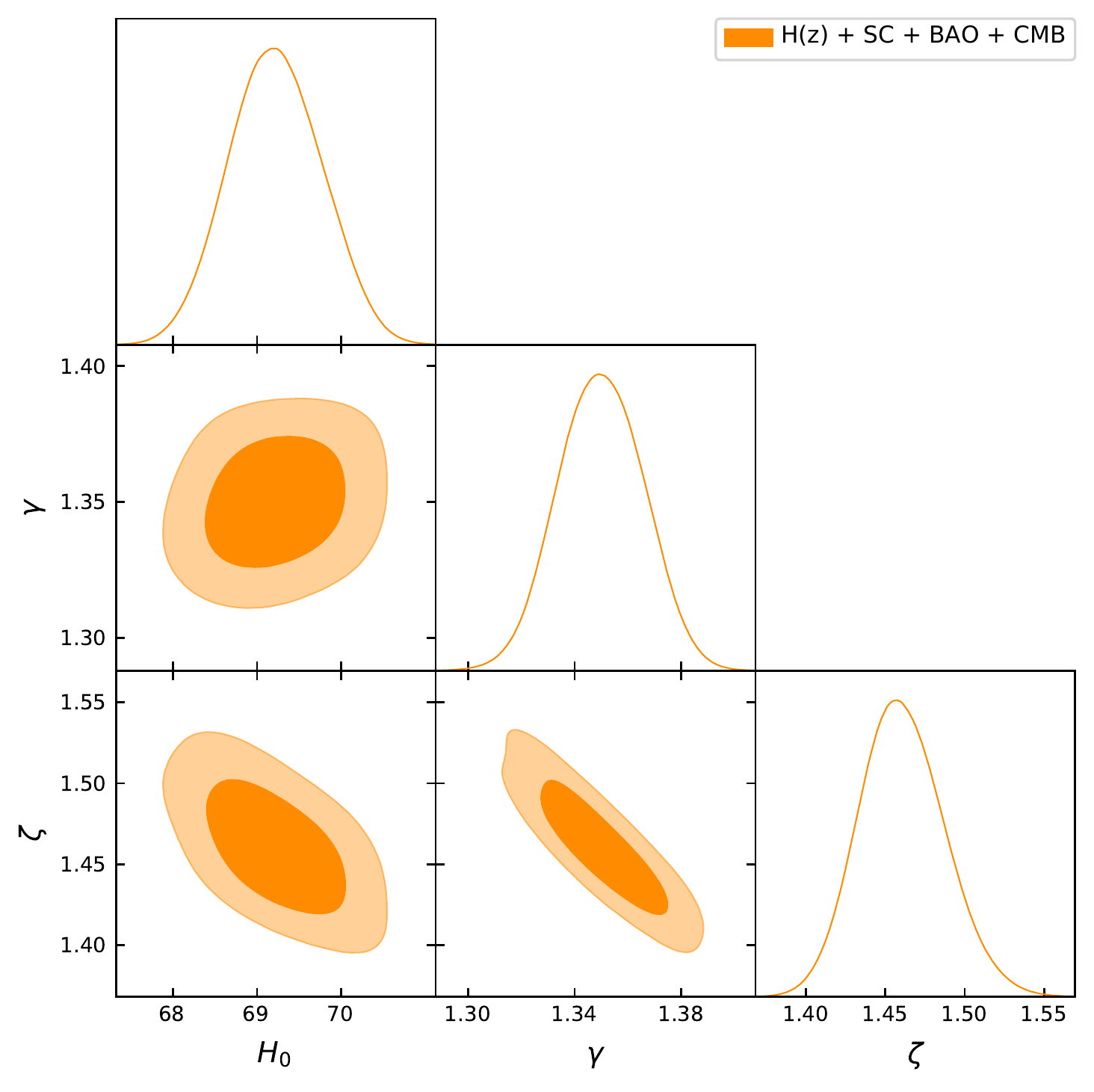}
\caption{MCMC confidence contours at 1$\protect\sigma$ and 2$\protect\sigma$ for Model 2.}\label{fig2}
   \end{minipage}
\end{figure}

\begin{table}[]
\begin{center}
\begin{tabular}{|c|c|c|}
\hline
\multicolumn{3}{|c|}{MCMC Results} \\ \hline
Model & Parameters & Bestfit Value \\ \hline
$\Lambda$CDM Model & $H_0$ & $69.854848_{-1.259100}^{+1.259100}$ \\ \hline
Model 1 & $H_0$ & $69.302413_{-0.443452}^{+0.443452}$ \\ 
& $\gamma$ & $1.662642_{-0.023615}^{+0.023615}$ \\ 
& $\zeta$ & $1.322252_{-0.027842}^{+0.027842}$ \\ \hline
Model 2 & $H_0$ & $69.247921_{-0.465229}^{+0.465229}$ \\ 
& $\gamma$ & $1.350387_{-0.015168}^{+0.015168}$ \\ 
& $\zeta$ & $1.459203_{-0.024204}^{+0.024204}$ \\ \hline
\end{tabular}%
\end{center}
\caption{Best fit values of the model parameters}\label{tab1}
\end{table}

\newpage
\section{Observational and theoretical comparisons of the Hubble Function and Distance Modulus Function}
After obtaining the best-fit values for the model parameters of Model 1 and 2, it is essential to compare these models with the widely accepted $\Lambda$CDM model. The $\Lambda$CDM model has demonstrated remarkable consistency with various observational datasets and is considered a robust framework for describing the evolution of the Universe. By comparing our parametrized models with the $\Lambda$CDM model, we can gain a deeper understanding of the deviations and discrepancies between the two. This comparative analysis allows us to investigate how our models differ from the $\Lambda$CDM model and explore the implications of these differences in the cosmological context. It provides insights into the specific aspects of our parametrized models that deviate from the $\Lambda$CDM Model, such as the expansion rate, matter content, and dynamics of the Universe. By examining the deviations between our models and the $\Lambda$CDM Model, we can identify the specific features and behaviors that distinguish our models. This analysis offers valuable information about the strengths and limitations of our parametrized models and provides insights into their potential implications for our understanding of the Universe. This comparison with the $\Lambda$CDM model serves as a benchmark for evaluating the viability and reliability of our models. It allows us to assess the goodness-of-fit of our models to observational data and determine the level of agreement between our parametrized models and the well-established $\Lambda$CDM framework.


\subsection{Comparison with the Hubble data points}
\begin{figure}[!htb]
   \begin{minipage}{0.49\textwidth}
     \centering
   \includegraphics[scale=0.6]{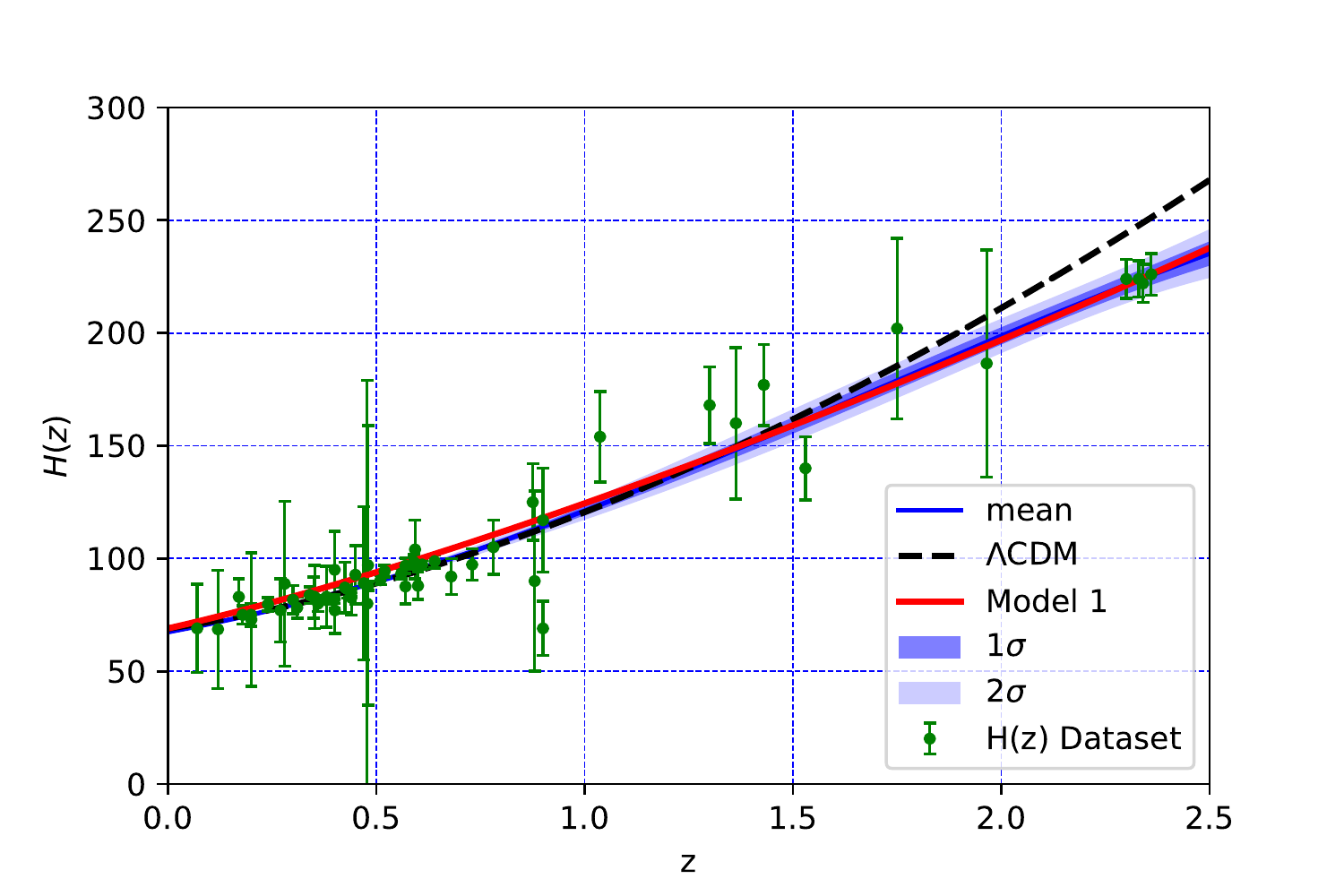}
\caption{The figure shows that the theoretical curve of the Hubble function $H(z)$ of Model 1 shown in red line and $\Lambda$CDM model shown in black dotted line with $\Omega_{\mathrm{m0}}=$ 0.3 and $\Omega_\Lambda =$ 0.7 against 57 $H(z)$ datasets are shown in green dots with their corresponding error bars with 1$\protect\sigma$ and 2$\protect\sigma$ error bands.}\label{H(z) Model 1}
   \end{minipage}\hfill
   \begin{minipage}{0.49\textwidth}
     \centering
    \includegraphics[scale=0.6]{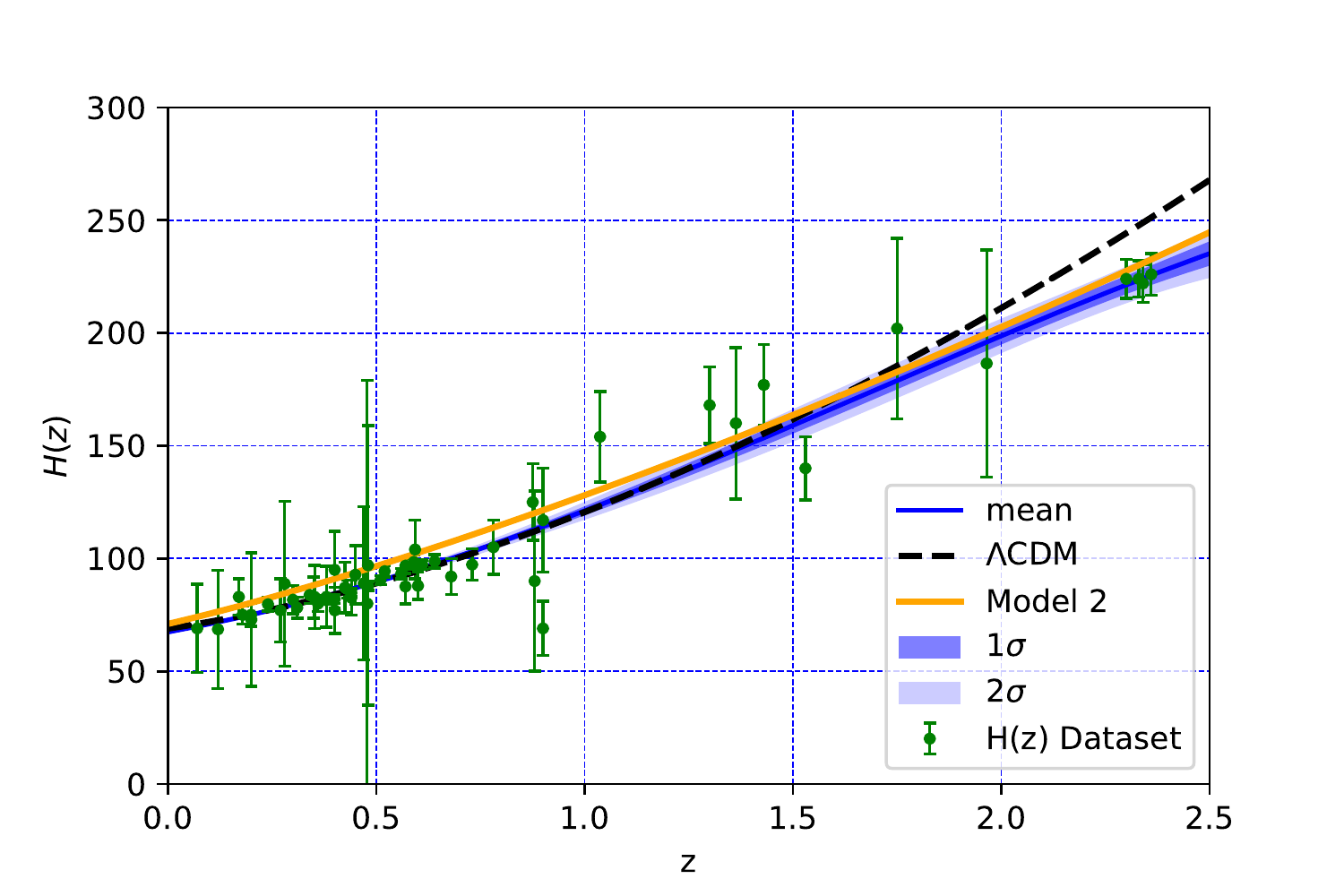}
\caption{The figure shows that the theoretical curve of the Hubble function $H(z)$ of Model 2 shown in orange line and $\Lambda$CDM model shown in black
dotted line with $\Omega_{\mathrm{m0}}=$ 0.3 and $\Omega_\Lambda =$ 0.7, against 57 $H(z)$ datasets are shown in green dots with their corresponding error bars with 1$\protect\sigma$ and 2$\protect\sigma$ error bands.}\label{H(z) Model 2}
   \end{minipage}
\end{figure}

In order to assess the agreement between our Model 1 and Model 2 with observational data, we compare their predictions to the Hubble data along with the $\Lambda$CDM model and its associated 1$\sigma$ and 2$\sigma$ error bands. The comparison is depicted in Fig:- \ref{H(z) Model 1} and Fig:- \ref{H(z) Model 2}. From these figures, it is evident that both Model 1 and Model 2 exhibit a good fit to the Hubble data. The data points align well with the predictions of our models, indicating that they capture the observed behavior of the Universe's expansion. Moreover, the agreement between our models and the Hubble data suggests that our parametrized frameworks offer plausible descriptions of cosmic evolution. The fact that Model 1 and 2 align well with the Hubble data provides support for their viability and indicates that they capture essential aspects of cosmic expansion. These findings demonstrate the satisfactory agreement of our models with the Hubble data, highlighting their potential to provide meaningful insights into the dynamics and evolution of the Universe within the framework of our proposed parametrized models.

\subsection{Comparison with the Pantheon dataset}
\begin{figure}[!htb]
   \begin{minipage}{0.49\textwidth}
     \centering
   \includegraphics[scale=0.6]{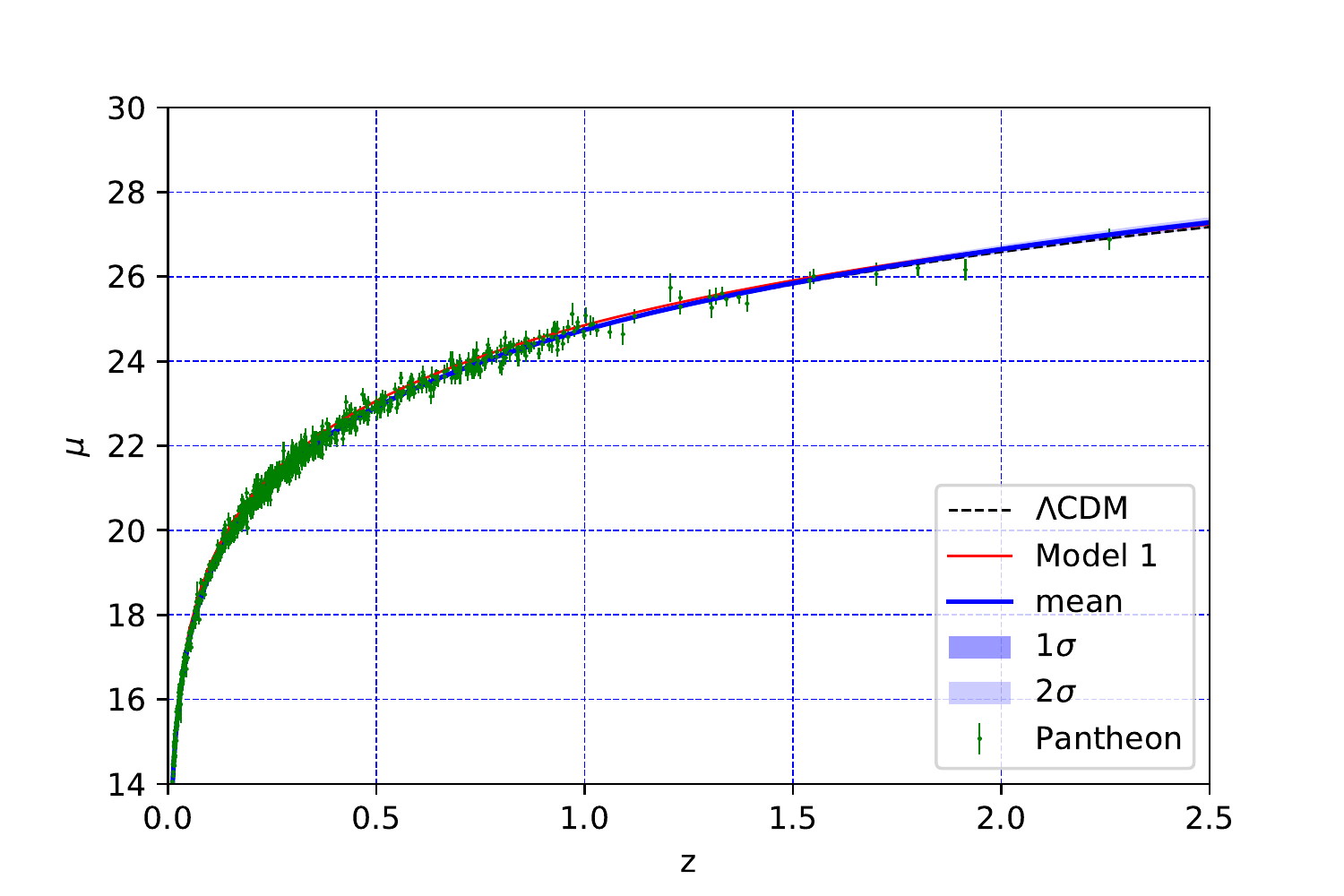}
\caption{Theoretical curve of distance modulus $\protect\mu(z) $ of the Model 1 is shown in red line and the $\Lambda$CDM model is shown in the black dotted line with $\Omega_{\mathrm{m0}}=$ 0.3 and $\Omega_\Lambda =$ 0.7, against type Ia supernova data are shown in green dots with their corresponding errors bars with 1$\protect\sigma$ and 2$\protect\sigma$ error bands.}\label{mu(z) Model 1}
   \end{minipage}\hfill
   \begin{minipage}{0.49\textwidth}
     \centering
    \includegraphics[scale=0.6]{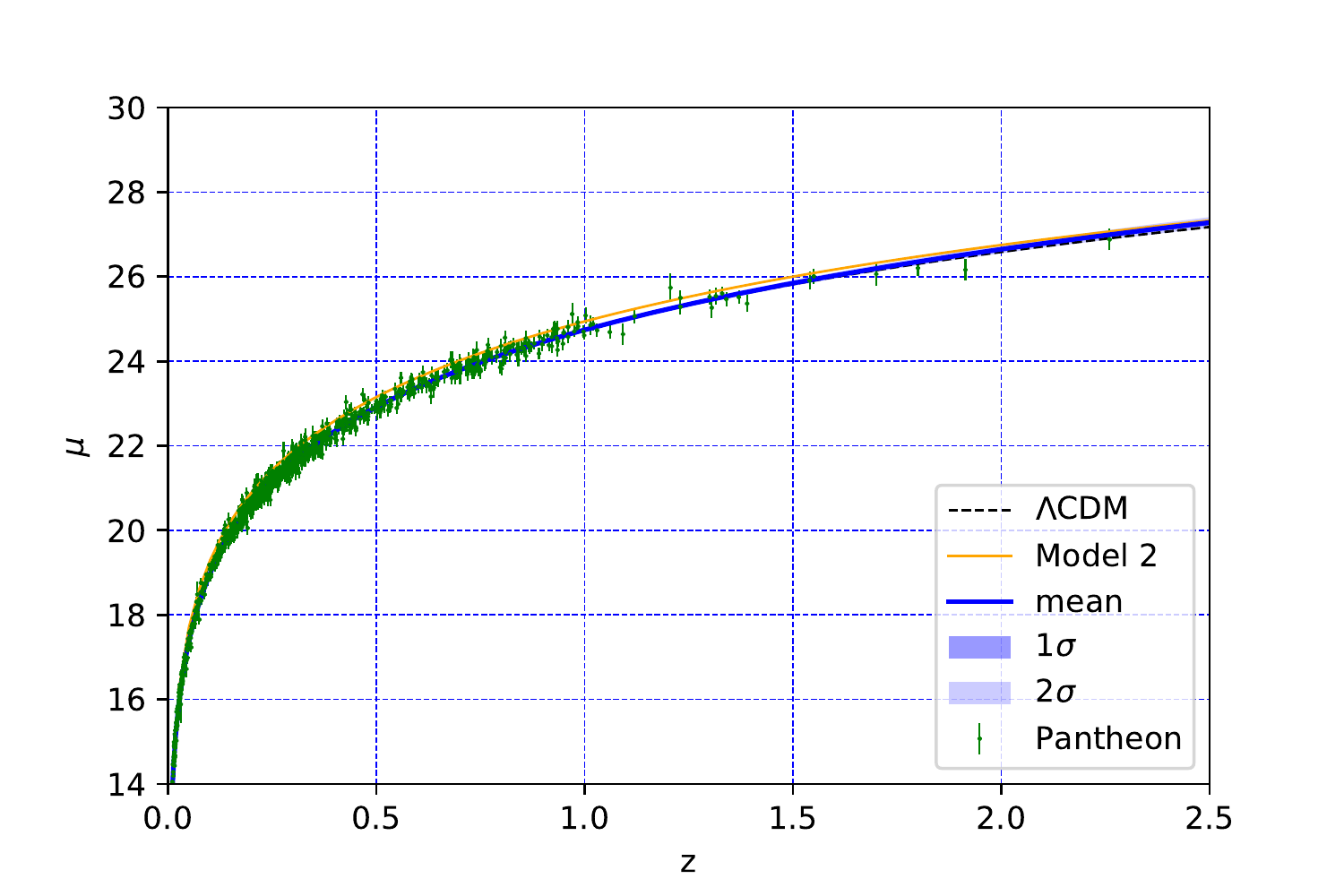}
\caption{Theoretical curve of distance modulus $\protect\mu(z) $ of the Model 2 shown in orange line and $\Lambda$CDM model shown in black dotted line with $\Omega_{\mathrm{m0}}=$ 0.3 and $\Omega_\Lambda =$ 0.7, against type Ia supernova data shown in green dots with their corresponding errors bars with 1$\protect\sigma$ and 2$\protect\sigma$ error bands.}
\label{mu(z) Model 2}
   \end{minipage}
\end{figure}

In this analysis, we compare the distance modulus function $\mu(z)$ of Model 1 and Model 2 with the Pantheon data, which consists of 1048 points and $\Lambda$CDM Model. The comparison is depicted in Fig:- \ref{mu(z) Model 1} and Fig:- \ref{mu(z) Model 2}. These figures demonstrate that both Model 1 and Model 2 provide a good fit to the Pantheon dataset and $\Lambda$CDM model, indicating that they are consistent with the observed distance measurements. This comparison with observational data provides support for the viability and reliability of our models in explaining the observed phenomena. It reinforces the notion that Model 1 and Model 2 are capable of reproducing the observed expansion history of the Universe as indicated by the Pantheon dataset.

\subsection{Relative difference between model and $\Lambda$CDM}

\begin{figure}[!htb]
   \begin{minipage}{0.49\textwidth}
     \centering
   \includegraphics[scale=0.43]{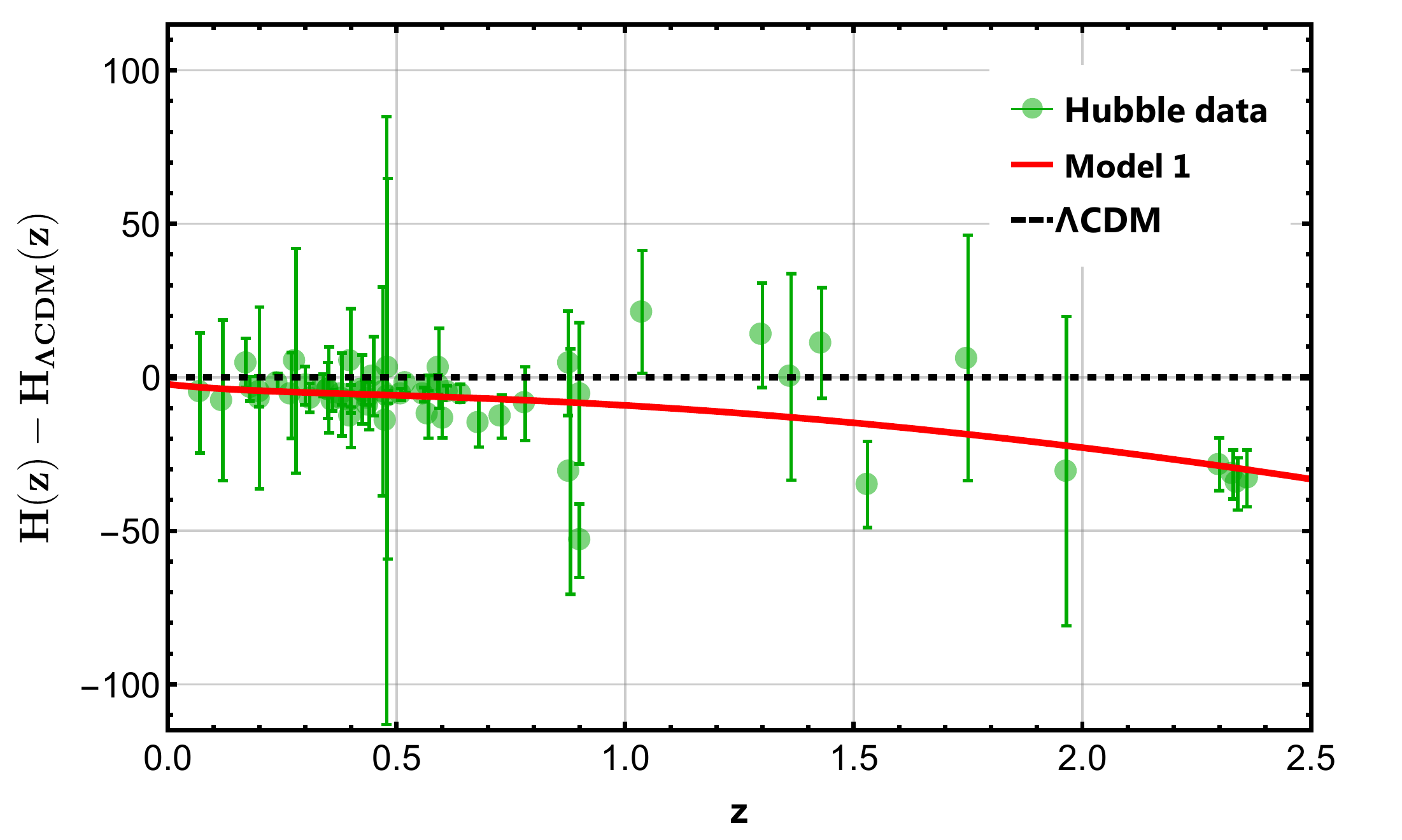}
\caption{The variation of the difference between Model 1 shown in the red line, and the $\Lambda$CDM model shown in the black dotted line with $\Omega_{\mathrm{m0}}=$ 0.3 and $\Omega_\Lambda =$ 0.7, as a function of the
redshift $z$ against the Hubble measurements, against 57 $H(z)$ datasets are shown in green dots with their corresponding error bars.}\label{h(z)diff1}
   \end{minipage}\hfill
   \begin{minipage}{0.49\textwidth}
     \centering
    \includegraphics[scale=0.43]{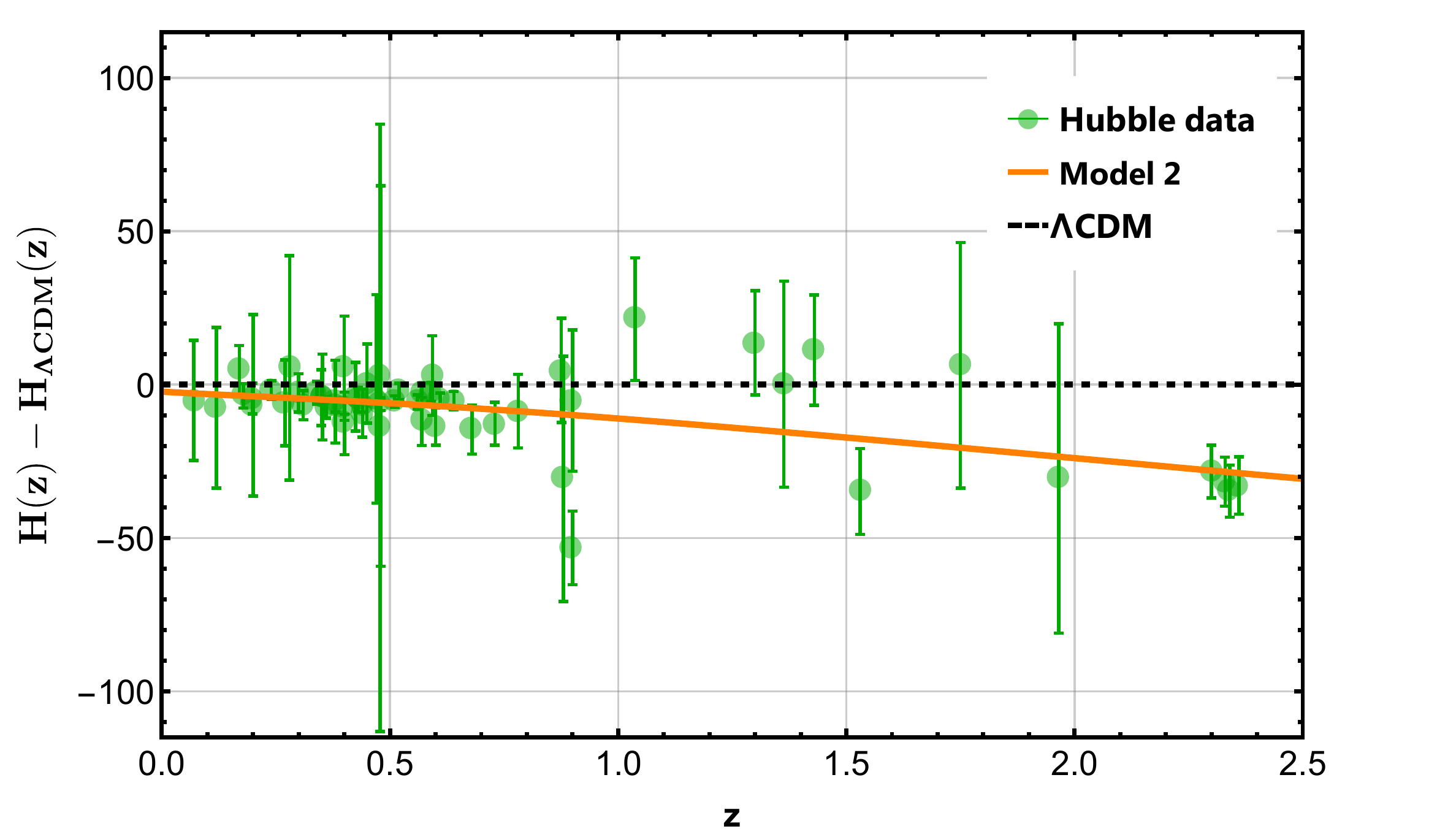}
\caption{The variation of the difference between Model 2 shown in the orange line, and the $\Lambda$CDM model shown in the black dotted line with $\Omega_{\mathrm{m0}}=$ 0.3 and $\Omega_\Lambda =$ 0.7, as a function of the redshift $z$ against the Hubble measurements, against 57 $H(z)$ datasets are shown in green dots with their corresponding error bars.}\label{h(z)diff2}
   \end{minipage}
\end{figure}

The relative difference between Model 1, Model 2 and the standard $\Lambda $%
CDM paradigm is shown in Fig:- \ref{h(z)diff1} and Fig:- \ref{h(z)diff2}. The
Figure demonstrates how the typical $\Lambda $CDM model and both models
perform very identically for $z<1$. However for $z>1$, there are some
discrepancies between both models and $\Lambda $CDM Model, and these
discrepancies do get greater as the redshift increases.

\section{Cosmic Evolution of Geometrical Parameters}
\subsection{The deceleration parameter}
The deceleration parameter (DP) is a dimensionless quantity that contributes
to the cosmological evaluation of the expansion rate. This parameter could
be expressed in terms of the scale factor, which decrease and causes the
cosmos to expand uniformly. Furthermore, negative values of this parameter
represent faster expansion, whereas positive values represent a decelerated
phase of the Universe. It is given mathematically by 
\begin{equation}
q=-\frac{a\ddot{a}}{\dot{a}^{2}}=-1+\frac{d}{dt}\left( \frac{1}{H}\right) 
\text{.}
\end{equation}%
Subsequently, one could notice that the Hubble parameter's value either
increases or decreases with time depending on the magnitude of DP. Different
ranges of the DP $q_{0}$ have been anticipated in various cosmological
scenarios. Ideally, the value of $q_{0}$ should be determined through
observational analysis. The DP, for example, is required for the connection
between apparent brightness and redshift for a class of identical supernovae
in distant galaxies though such estimates are difficult to understand. The
latest findings completely corroborate the accelerating Universe
speculations. One has to be extremely accurate in determining the value of $%
q_{0}$. It turned out to be highly model-dependent, and the evidence for an
accelerating Universe is not as convincing as commonly supposed. For Model 1
and 2, the expression for the $q(z)$ is given by

\begin{equation}
q(z)=-1+\gamma -2\gamma \left[ 1+\{\zeta (1+z)\}^{\gamma }\right] ^{-1}\text{%
.}  \label{30}
\end{equation}

\begin{equation}
q(z)=-1+\gamma -3\gamma \left[ 1+\{\zeta (1+z)\}^{2\gamma }\right] ^{-1}%
\text{.}  \label{31}
\end{equation}

\begin{figure}[!htb]
   \begin{minipage}{0.49\textwidth}
     \centering
   \includegraphics[scale=0.43]{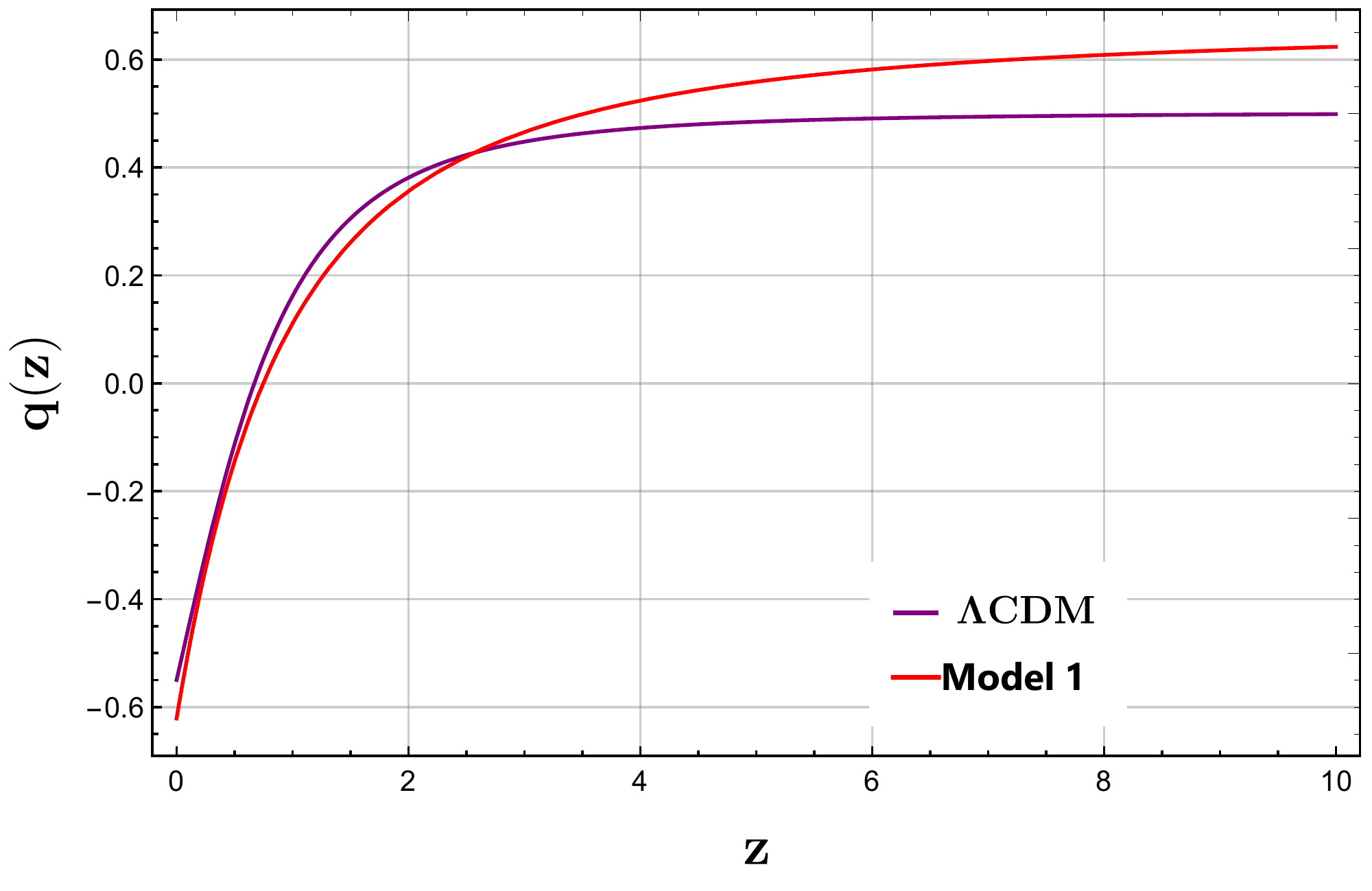}
\caption{Evolution of deceleration parameter with respect to the redshift of Model 1.}\label{q(z) Model 1}
   \end{minipage}\hfill
   \begin{minipage}{0.49\textwidth}
     \centering
    \includegraphics[scale=0.43]{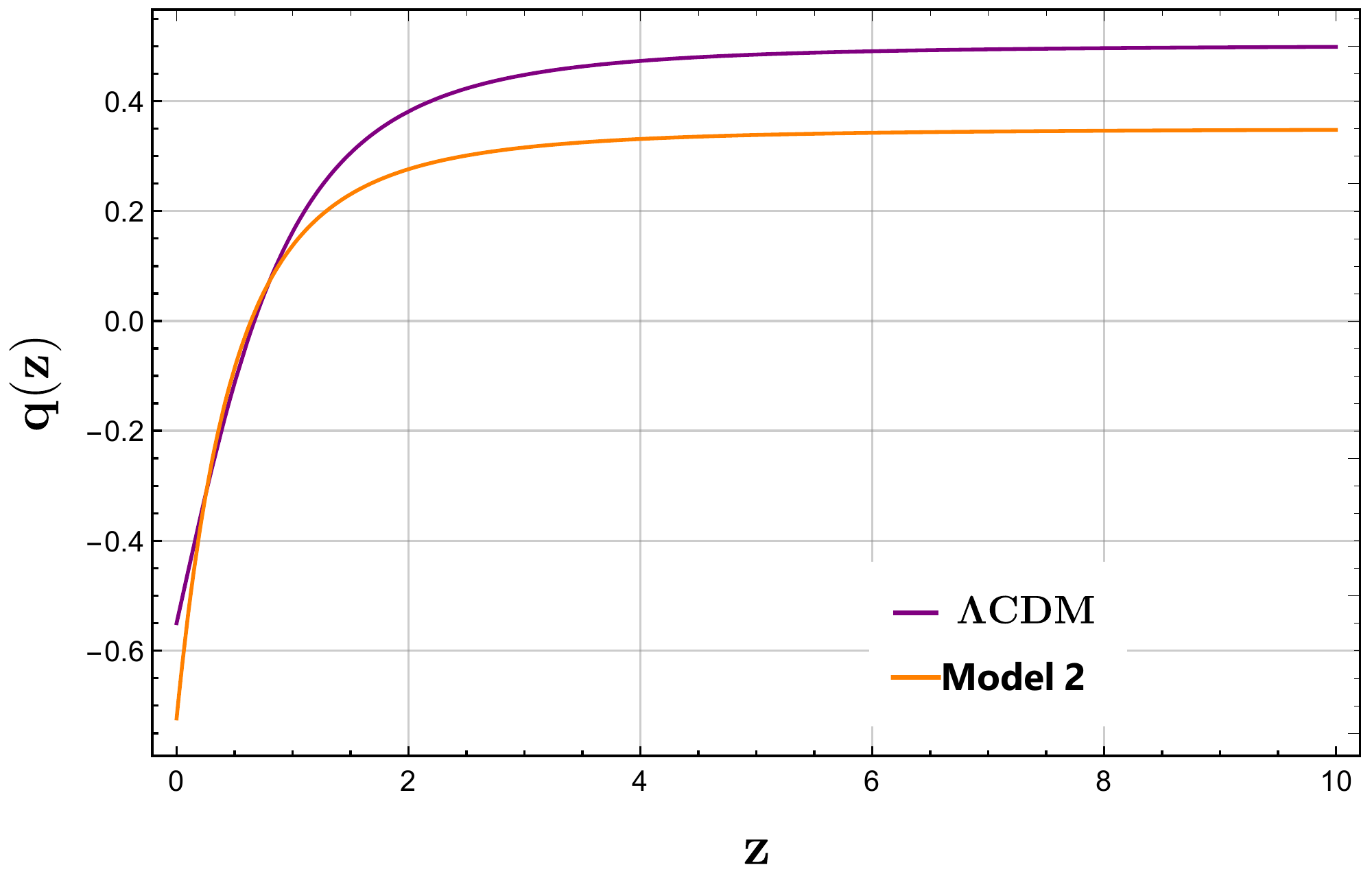}
\caption{Evolution of deceleration parameter with respect to the redshift of Model 2.}\label{q(z) Model 2}
   \end{minipage}
\end{figure}

The redshift evolution of the deceleration parameter for the models are shown in the figures in Fig:- \ref{q(z) Model 1} and Fig:- \ref{q(z) Model 2}.  

\subsection{The jerk parameter}

The dimensionless jerk parameter (popularly referred to as jolt) is a
refinement of the standard cosmological parameters $a(t)$ and $q(t)$. Other
synonyms for jerk include impulse, bounce, surge, shock, and
super-acceleration. The jerk parameter could potentially be represented in
terms of the third-order derivative of the scale factor with respect to
cosmic time, yielding an absolute approach to abandoning the concordance $%
\Lambda $CDM model. We may express it mathematically as,

\begin{equation}
j=\frac{1}{a}\frac{d^{3}a}{d\tau ^{3}}\left[ \frac{1}{a}\frac{da}{d\tau }%
\right] ^{-3}=q(2q+1)+(1+z)\frac{dq}{dz}\text{.}  \label{32}
\end{equation}

\begin{figure}[!htb]
   \begin{minipage}{0.49\textwidth}
     \centering
   \includegraphics[scale=0.42]{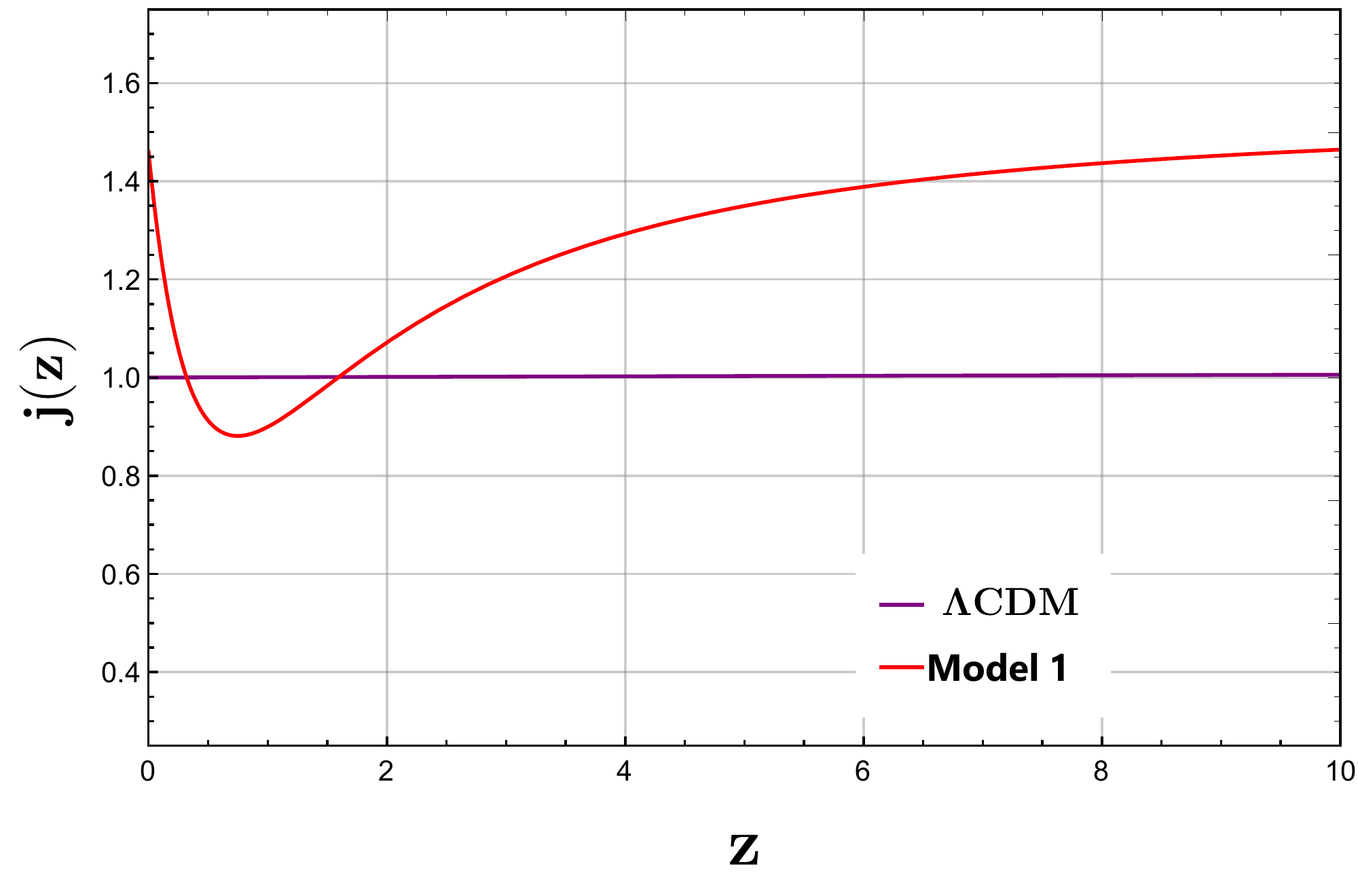}
\caption{Evolution of jerk parameter with respect to the redshift of Model 1.}\label{j(z) Model 1}
   \end{minipage}\hfill
   \begin{minipage}{0.49\textwidth}
     \centering
    \includegraphics[scale=0.42]{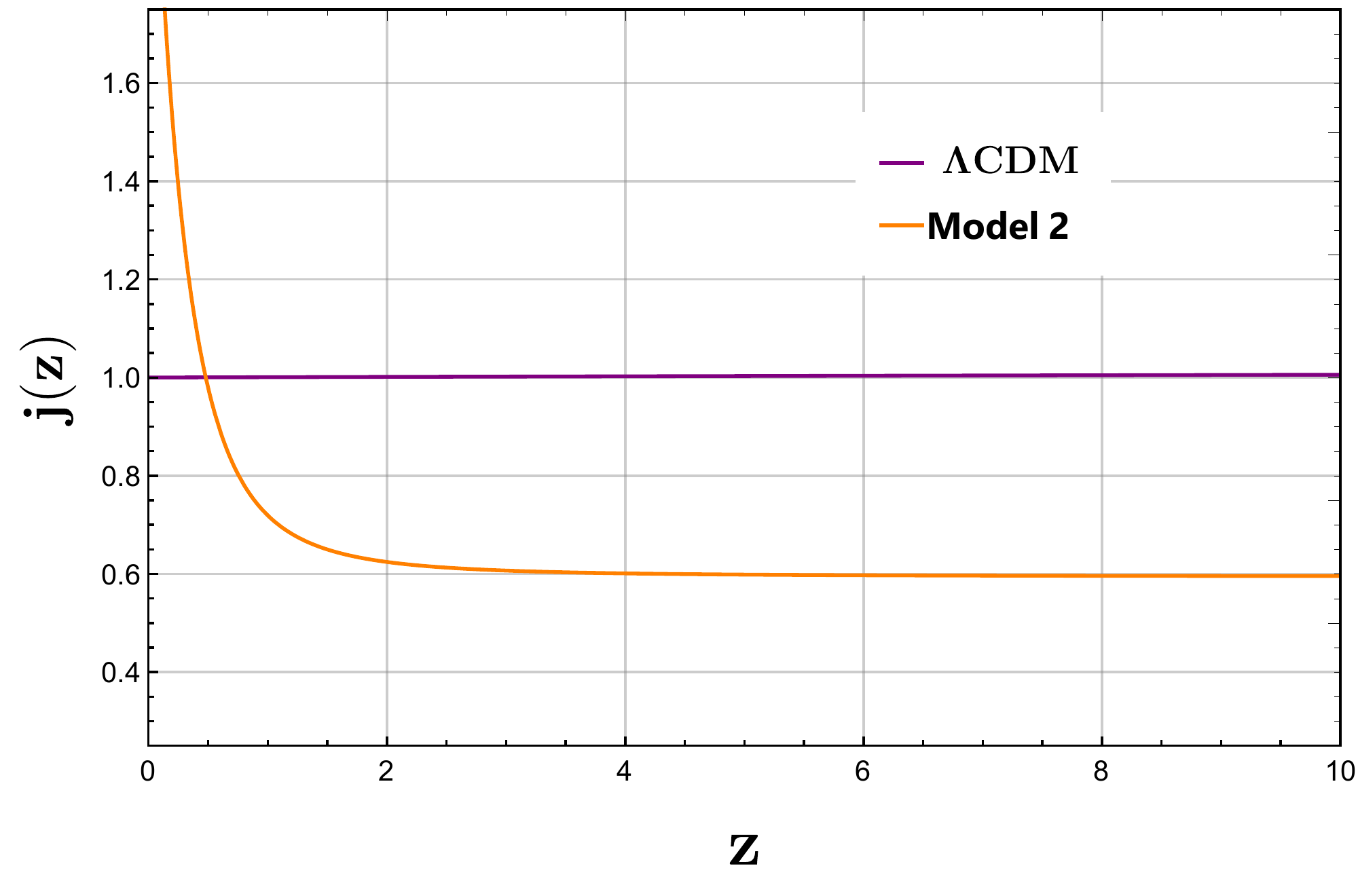}
\caption{Evolution of jerk parameter with respect to the redshift of Model 2.}\label{j(z) Model 2}
   \end{minipage}
\end{figure}

The following plots, in the figures, Fig:- \ref{j(z) Model 1} and Fig:- \ref{j(z) Model 2} show the evolution of the jerk parameter for both Model 1 and Model 2.  

\subsection{Snap parameter}

The Snap parameter (the fourth time derivative) is also known as jounce. The
fifth and sixth-time derivatives are frequently jokingly referred to as
crackle and pop. The dimensionless snap parameter is defined as follows:

\begin{equation}
s=\frac{1}{a}\frac{d^{4}a}{d\tau ^{4}}\left[ \frac{1}{a}\frac{da}{d\tau }%
\right] ^{-4}=\frac{j-1}{3\left( q-\frac{1}{2}\right) },  \label{33}
\end{equation}

\begin{figure}[!htb]
   \begin{minipage}{0.49\textwidth}
     \centering
   \includegraphics[scale=0.42]{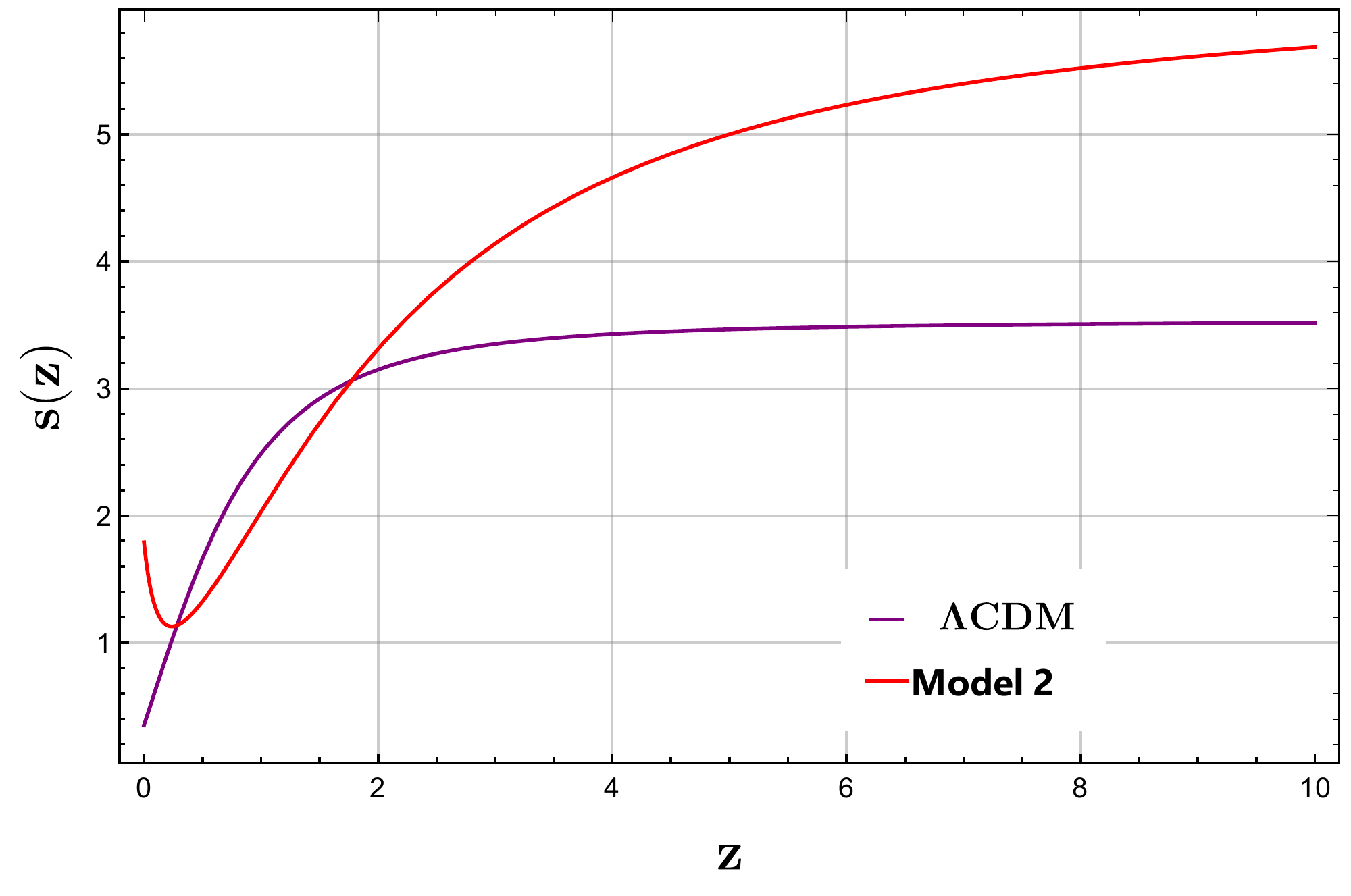}
\caption{Evolution of snap parameter with respect to the redshift of Model 1.}\label{s(z) Model 1}
   \end{minipage}\hfill
   \begin{minipage}{0.49\textwidth}
     \centering
    \includegraphics[scale=0.42]{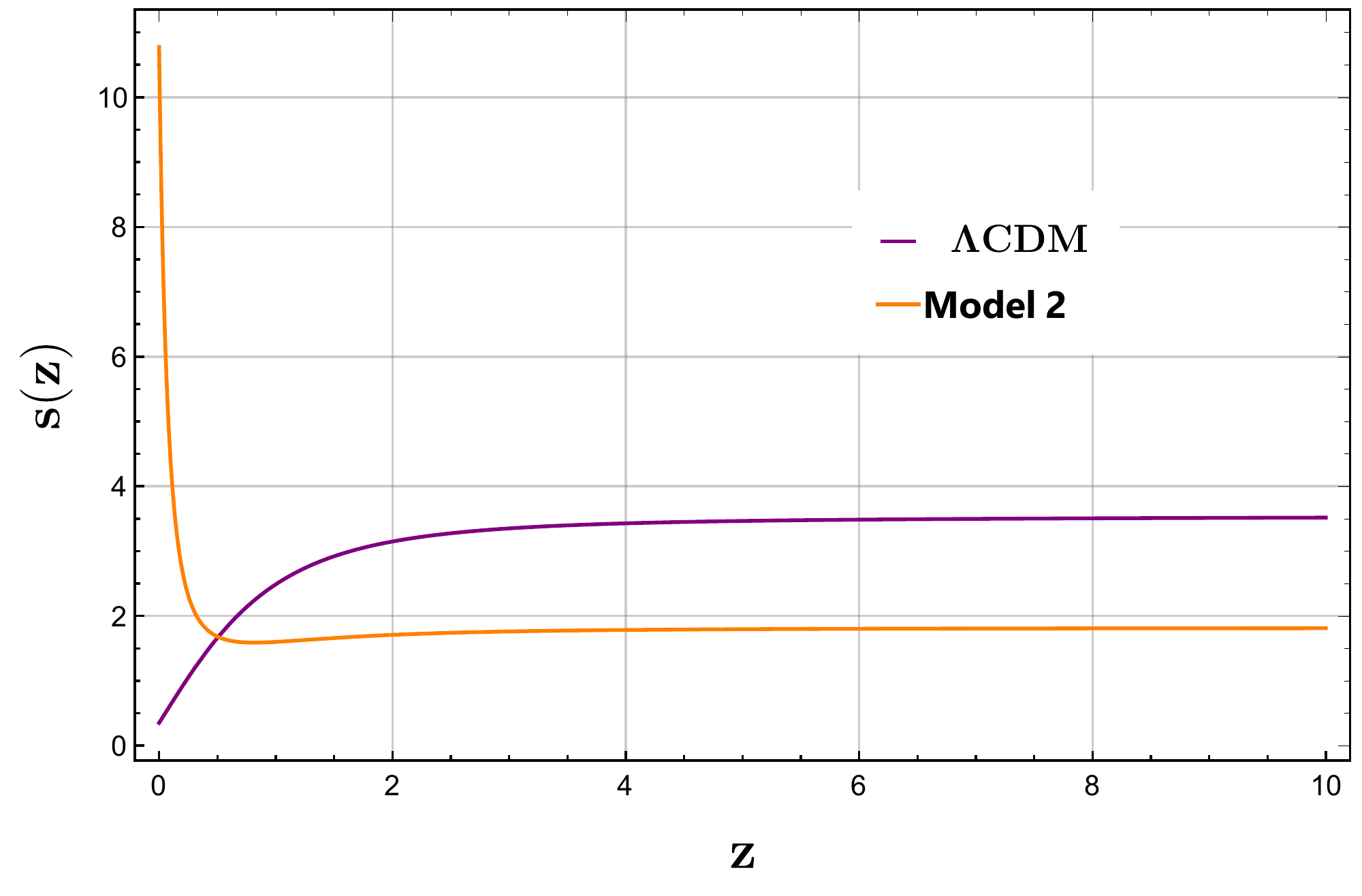}
\caption{Evolution of snap parameter with respect to the redshift of Model 2.}\label{s(z) Model 2}
   \end{minipage}
\end{figure}

The following plots, in the figures, Fig:- \ref{s(z) Model 1} and Fig:- \ref{s(z) Model 2} show the evolution of the snap parameter for both Model 1 and Model 2.

\section{Diagnostic Analysis of the models}

\subsection{Statefinder diagnostic}

When interpreting various cosmological circumstances containing DE, an
effective and significant diagnostic of DE is needed. Sahni et al.\cite{76,
77} proposed a novel DE diagnosis method based on higher derivatives of the
scale factor. Statefinder diagnostics \cite{76, 77, 78, 79} is a technique
that is commonly used to distinguish and contrast the characteristics of
multiple DE models utilizing higher-order derivatives of the scale factor.
The cosmological statefinder diagnostic pair $\{r,s\}$ permits one to
examine the cosmic properties of DE in a model-independent way, which may be
determined through relationships, 
\begin{equation}
r=\frac{\dddot{a}}{aH^{3}},s=\frac{r-1}{3\left( q-\frac{1}{2}\right) }\text{.%
}
\end{equation}%
In this case, the parameter $s$ is a linear amalgamation of $r$ and $q$.
Since it is calculated in terms of the cosmic scale factor, this diagnostic
pair is dimensionless and geometrical. Various possibilities in the $\{r,s\}$
and $\{q,r\}$ planes are exhibited to depict the temporal evolution of
various DE models. With the assistance of the statefinder diagnostics pair.
In these cases, some specific pairs typically correlate to classic DE models
such as $\{r,s\}=\{1,0\}$ represents $\Lambda $CDM model and $\{r,s\}=\{1,1\}
$ indicates standard cold dark matter Model (SCDM) in FLRW background. Also, 
$(-\infty ,\infty )$ yields static Einstein Universe. In the $r-s$ plane, $%
s>0$ and $s<0$ define a quintessence-like model and phantom-like model of
the DE, respectively. Moreover, the evolution from phantom to quintessence
can be observed by deviation from ${r,s}={1,0}$. On the other hand, $%
\{q,r\}=\{-1,1\}$ corresponds to the $\Lambda \mathrm{CDM}$ model while $%
\{q,r\}=\{0.5,1\}$ shows SCDM model. It is important to note that on a $r-s$
plane if the DE model's trajectories deviate from these standard values, the
resulting model differs from the normal cosmic models. For Model 1

\begin{equation}
\begin{aligned} r(z) = 1+\gamma(2 \gamma-3)+\frac{6
\gamma}{1+\{\zeta(1+z)\}^\gamma} \left[1-\gamma
\frac{\gamma}{1+\{\zeta(1+z)\}^\gamma}\right] \\. \end{aligned}  \label{38}
\end{equation}

\begin{equation}
\begin{aligned} s(z) = \frac{2
\gamma}{3}-\frac{\gamma}{1+\{\zeta(1+z)\}^\gamma}+\frac{\gamma(3+2
\gamma)}{3\left[-3-2 \gamma+(2 \gamma-3)\{\zeta(1+z)\}^\gamma\right]} \\.
\end{aligned}  \label{39}
\end{equation}

for Model 2

\begin{equation}
\begin{aligned} r(z)& =1-3 \gamma+2 \gamma^2+\frac{12
\gamma^2}{\left[1+\{\zeta(1+z)\}^{2 \gamma}\right]^2}+\frac{3
\gamma(3-2 \gamma)}{1+\{\zeta(1+z)\}^{2 \gamma}} \end{aligned}  \label{40}
\end{equation}

\begin{equation}
\begin{aligned} s(z)& = \frac{2}{3}\gamma-\frac{4
\gamma}{3+3\{\zeta(1+z)\}^{2 \gamma}}+\frac{2\gamma(3+4
\gamma)}{\left[-9-12 \gamma+3(2 \gamma-3)\{\zeta(1+z)\}^{2 \gamma}\right]}
\\. \end{aligned}  \label{41}
\end{equation}

The figures presented below, Fig:- \ref{rs1}, Fig:- \ref{rq1} and Fig:- \ref{rs2}, Fig:- \ref{rq2}, depict the dynamic changes in the these statefinder parameters for both Model 1 and Model 2.

\begin{figure}[!htb]
   \begin{minipage}{0.49\textwidth}
     \centering
   \includegraphics[scale=0.42]{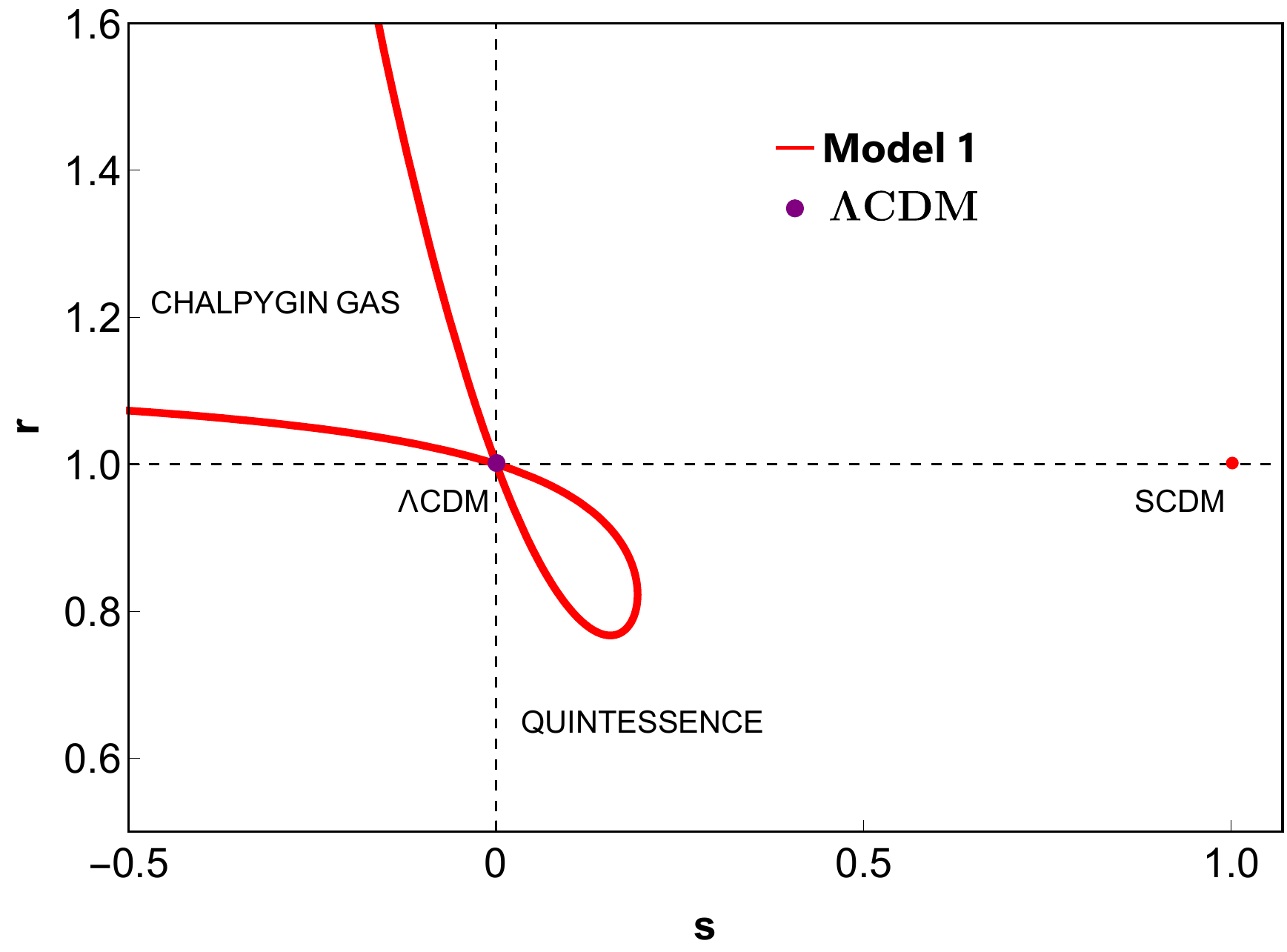}
\caption{This figure shows $\{s, r\}$ plots for Model 1.}\label{rs1}
   \end{minipage}\hfill
   \begin{minipage}{0.49\textwidth}
     \centering
    \includegraphics[scale=0.42]{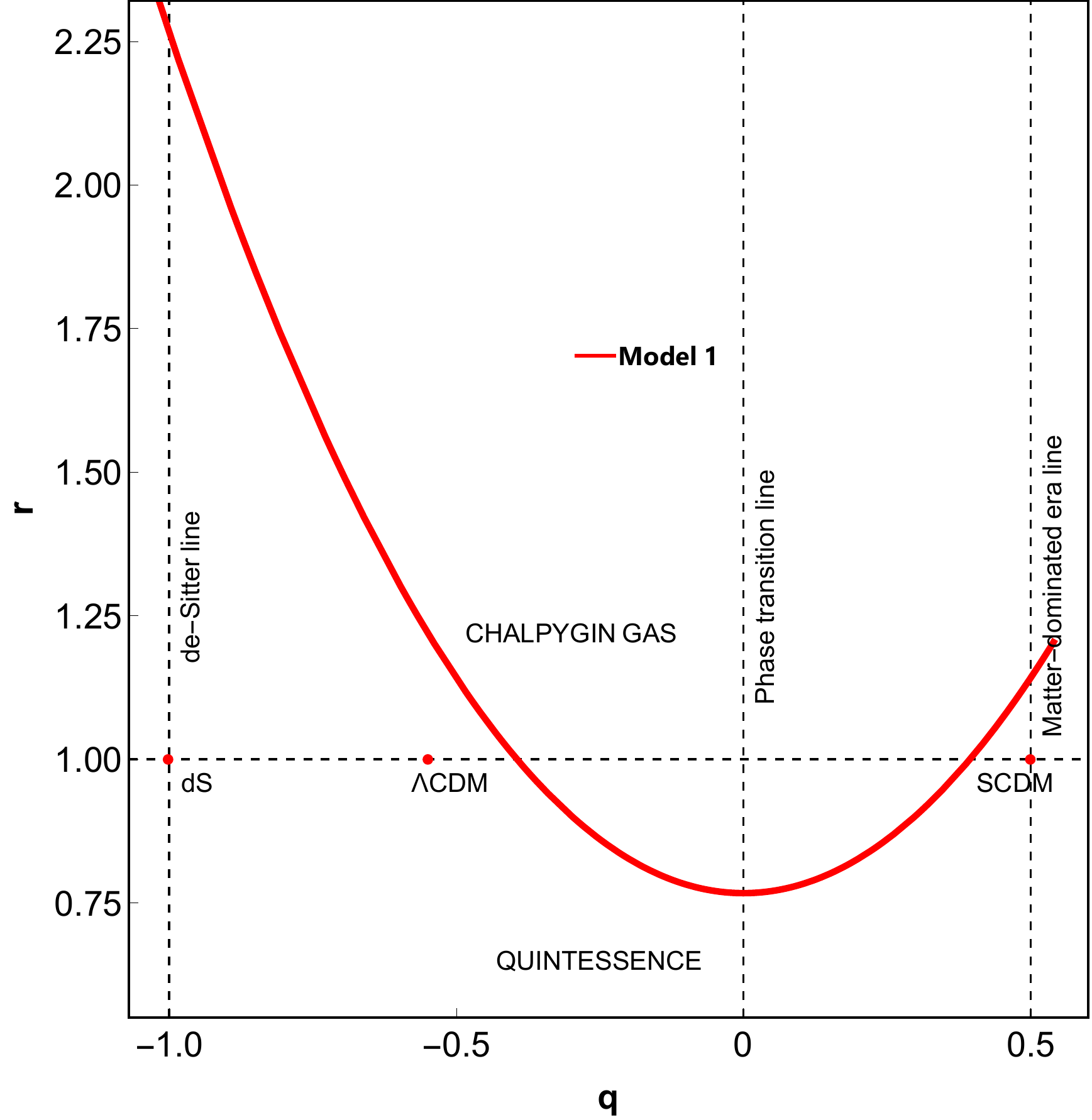}
\caption{This figure shows $\{q, r\}$ plots for Model 1.}
\label{rq1}
   \end{minipage}
\end{figure}
\begin{figure}[!htb]
   \begin{minipage}{0.49\textwidth}
     \centering
   \includegraphics[scale=0.42]{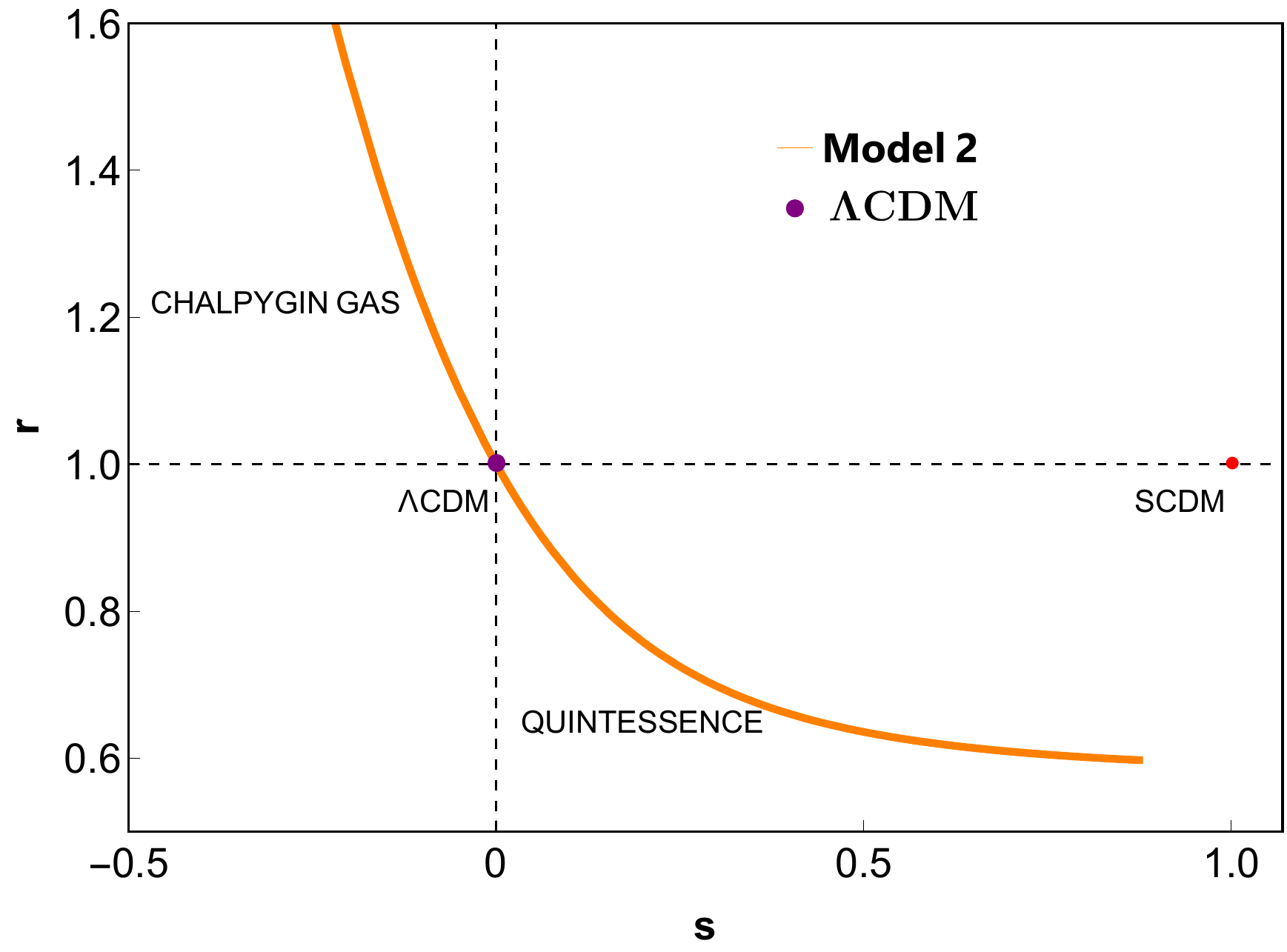}
\caption{This figure shows $\{s, r\}$ plots for Model 2.}\label{rs2}
   \end{minipage}\hfill
   \begin{minipage}{0.49\textwidth}
     \centering
    \includegraphics[scale=0.42]{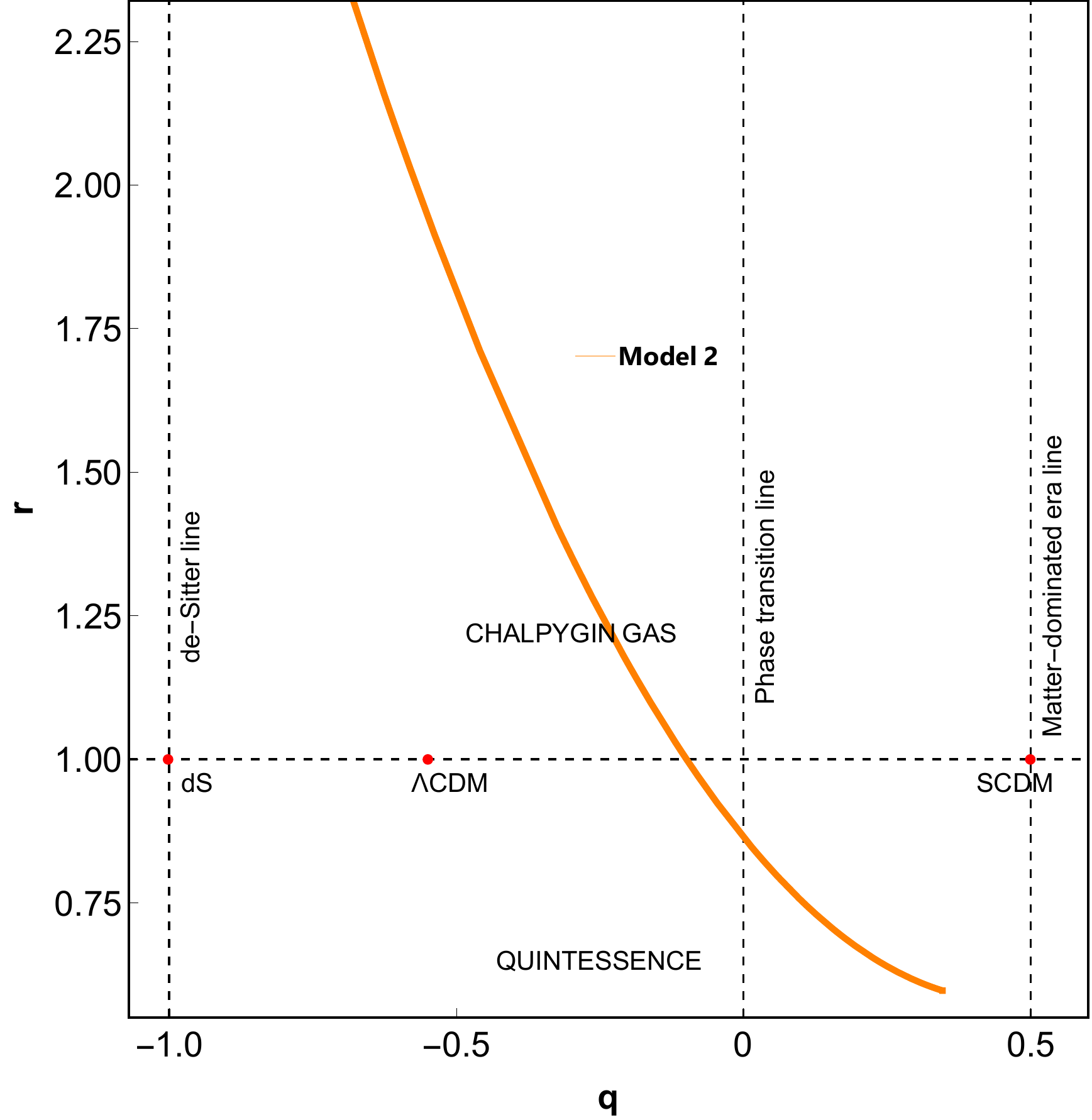}
\caption{This figure shows $\{q, r\}$ plots for Model 2.}
\label{rq2}
   \end{minipage}
\end{figure}

\subsection{Om Diagnostic}
Om diagnostic \cite{80, 81, 82, 83} is a geometrical study that uses the Hubble parameter to establish a null test for the $\Lambda $CDM model. Similarly to the statefinder diagnostic, the Om diagnostic efficiently
separates distinct DE models from $\Lambda $CDM by varying the slope of $Om(z)$. A positive slope of the diagnostic parameter represents a quintessence model, whereas a negative slope represents a phantom model. A constant slope with respect to redshift also determines the nature of DE, which coincides with the cosmological constant. $Om(z)$ is defined in the situation of a flat Universe by,

\begin{equation}
Om(z)=\frac{\left( \frac{H(z)}{H_{0}}\right) ^{2}-1}{(1+z)^{3}-1}\text{.}
\label{34}
\end{equation}%
One could express $Om(z)$ expression for Model 1 as, 
\begin{equation}
Om(z)=\frac{\frac{\left[ 1+\{\zeta (1+z)\}^{\gamma }\right\} ^{4}}{\left(
1+\zeta ^{\gamma }\right) ^{4}(1+z)^{2\gamma }}-1}{(1+z)^{3}-1}\text{.}
\label{35}
\end{equation}%
For Model 2 express of $Om(z)$ read as, 
\begin{equation}
Om(z)=\frac{\frac{\left[ 1+\{\zeta (1+z)\}^{2\gamma }\right] ^{3}}{\left(
1+\zeta ^{2\gamma }\right) ^{3}(1+z)^{4\gamma }}-1}{(1+z)^{3}-1}\text{.}
\label{36}
\end{equation}%

\begin{figure}[!htb]
   \begin{minipage}{0.49\textwidth}
     \centering
   \includegraphics[scale=0.42]{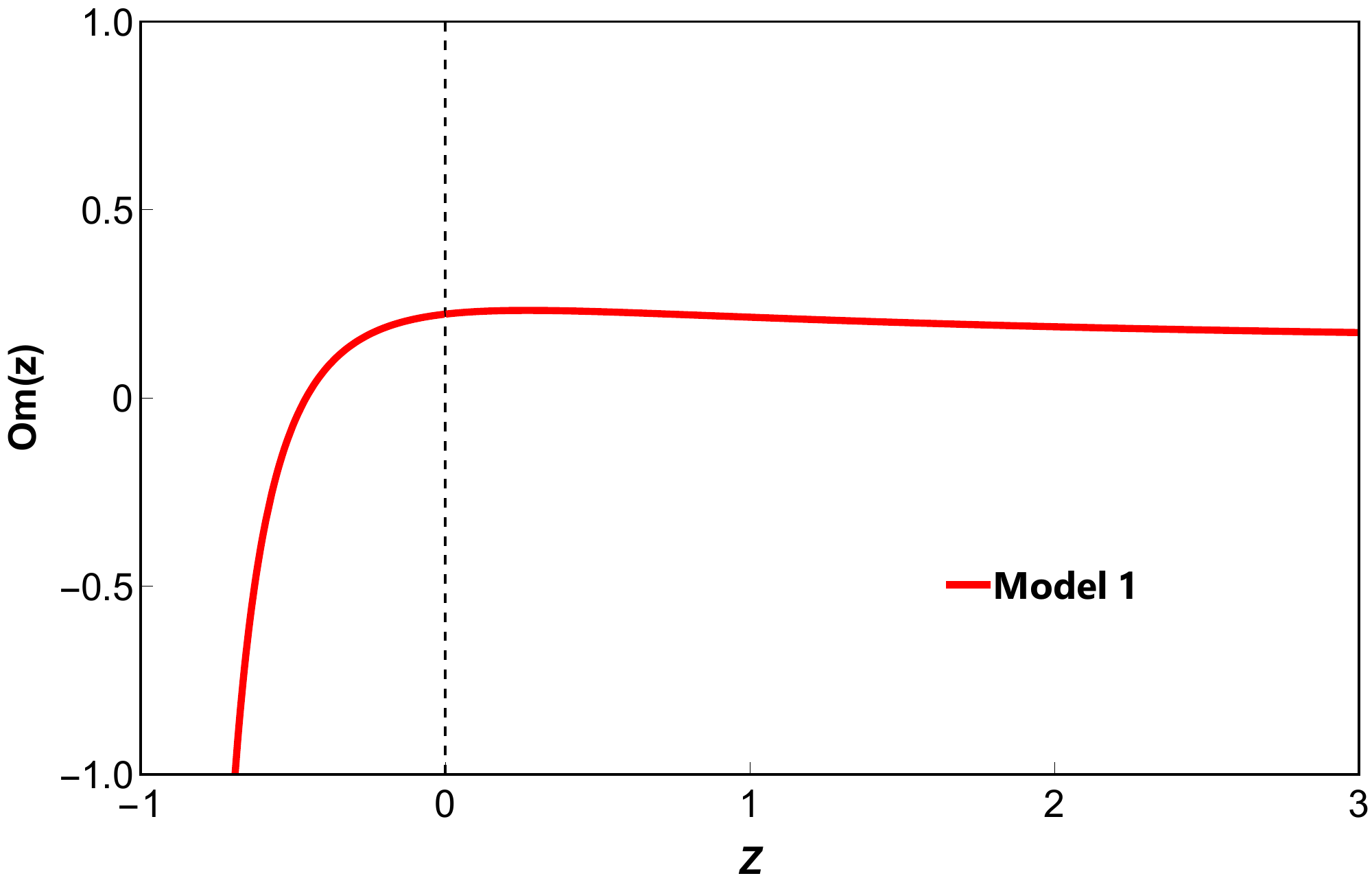}
\caption{This figure shows the $Om(z)$ with respect to redshift for Model 1.}\label{Om1}
   \end{minipage}\hfill
   \begin{minipage}{0.49\textwidth}
     \centering
    \includegraphics[scale=0.42]{Omz_1.pdf}
\caption{This figure shows the $Om(z)$ with respect to redshift for Model 2.}\label{Om2}
   \end{minipage}
\end{figure}

Fig:- \ref{Om1} and Fig:- \ref{Om2} depict the evolution of $\mathrm{Om}(z)$ with the redshift $z$ for Model 1 and Model 2, respectively.

\section{Cosmic Evolution of Physical Parameters for quintessence as a source of DE}

For quintessence as a candidate of dark energy, we have from equations (\ref%
{eq5}) and (\ref{eq6}), the expressions for the quintessence energy and pressure can be expressed as, 
\begin{equation}
M_{Pl}^{-2}\rho _{\phi }=3H^{2}-M_{Pl}^{-2}\rho _{M},  \label{s1}
\end{equation}%
\begin{equation}
M_{Pl}^{-2}p_{\phi }=(2q-1)H^{2},  \label{s2}
\end{equation}%
with the understanding of negligible pressure due to dust matter ($p_{M}=0$%
). For a two fluid Universe, scalar field, and matter, we have the minimal
interaction between Matter \& DE. When there is minimal interaction between
the matter component and the dark energy, they conserve separately for
which, we have $\dot{\rho}_{M}+3H\rho _{M}=0$ and $\dot{\rho}_{\phi }+3H\rho
_{\phi }=0$. This yield, $\rho _{M}=ca^{-3}=c(1+z)^{3}$, $c$ is a constant
of integration. At $t=t_{0}$ ($z=0$) and in terms of the density parameter ($%
\Omega $), we have, $c=3M_{Pl}^{-2}H_{0}^{-2}\Omega _{M0}$, which implies $%
\rho _{M}=3M_{Pl}^{-2}H_{0}^{-2}\Omega _{M0}(1+z)^{3}$. Here and afterward,
the suffix $0$ stands for the values of the cosmological parameters at
present time ($t=t_{0}$ or $z=0$).

\subsection{Energy Density \& Pressure of DE}

Now, solving, equations. (\ref{s1}) and (%
\ref{s2}), we obtain the expressions of energy density and pressure for the quintessence field for Model 1 as,

\begin{equation}
M_{Pl}^{-2}H_{0}^{-2}\rho _{\phi } = 3[ \left( 1+\zeta
^{\gamma }\right)^{-4}(1+z)^{-2\gamma }\left[ 1+\{\zeta (1+z)\}^{\gamma}\right] ^{4} -\Omega_{M0}(1+z)^{3}] 
\end{equation}

\begin{equation}
M_{Pl}^{-2}H_{0}^{-2}p_{\phi } = \left\{ -3+2\gamma
-4\gamma \left[ 1+\{\zeta (1+z)\}^{\gamma }\right] ^{-1}\right\}\left\{ \left( 1+\zeta ^{\gamma}\right) ^{-4}(1+z)^{-2}\left[ 1+\{\zeta
(1+z)\}^{\gamma }\right]^{4}\right\}.
\end{equation}

In order to observe the past, the present and future evolution of these physical parameters clearly, we have plotted them w.r.t. redshift '$z$' and shown them graphically in the following Fig:- \ref{fig:densitym1} \& Fig:- \ref{fig:pressurem1} for Model 1 with the found constrained values of the model parameters.

\begin{figure}[!htb]
   \begin{minipage}{0.49\textwidth}
     \centering
   \includegraphics[scale=0.42]{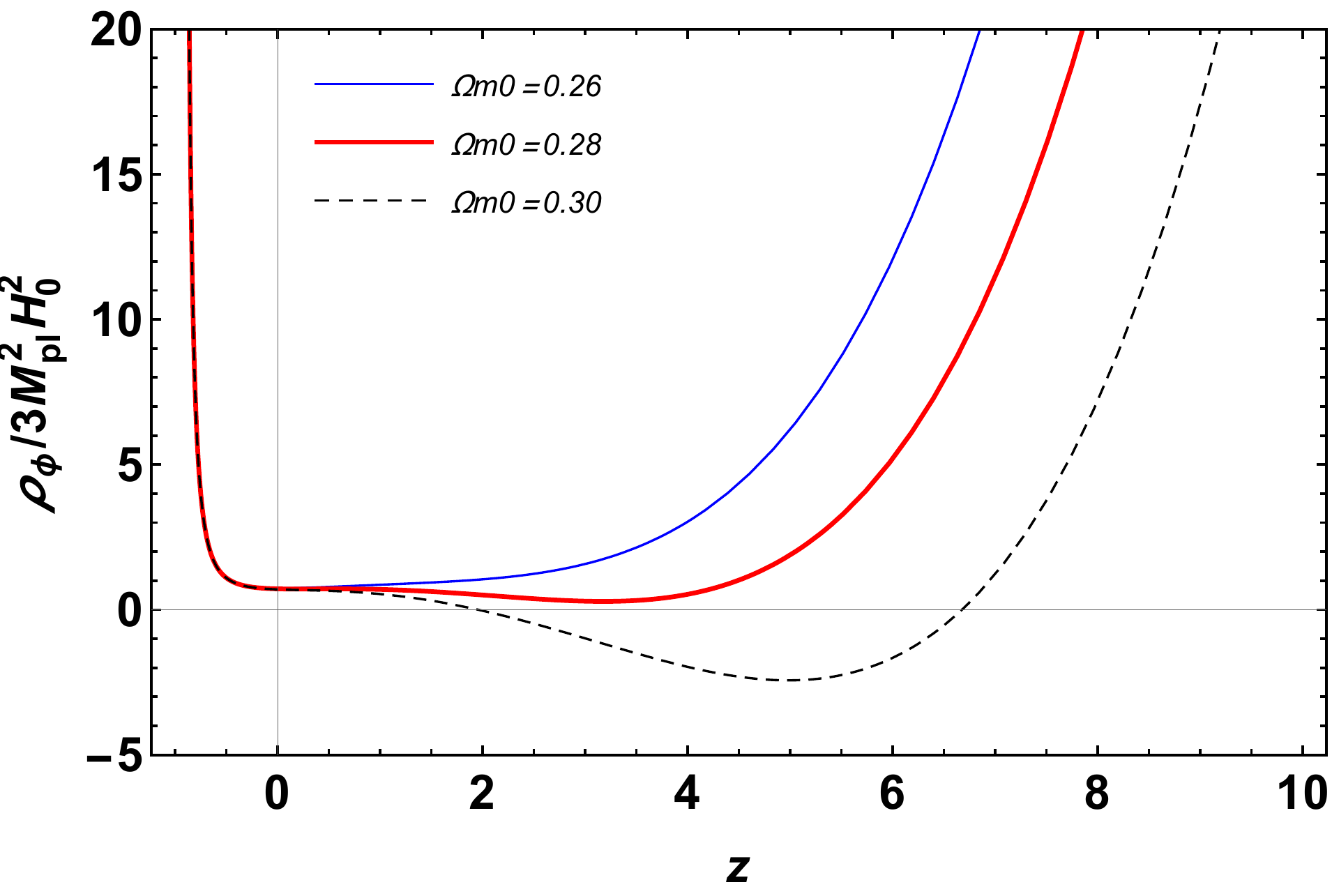}
\caption{Profile of energy density for Model 1.}\label{fig:densitym1}
   \end{minipage}\hfill
   \begin{minipage}{0.49\textwidth}
     \centering
    \includegraphics[scale=0.42]{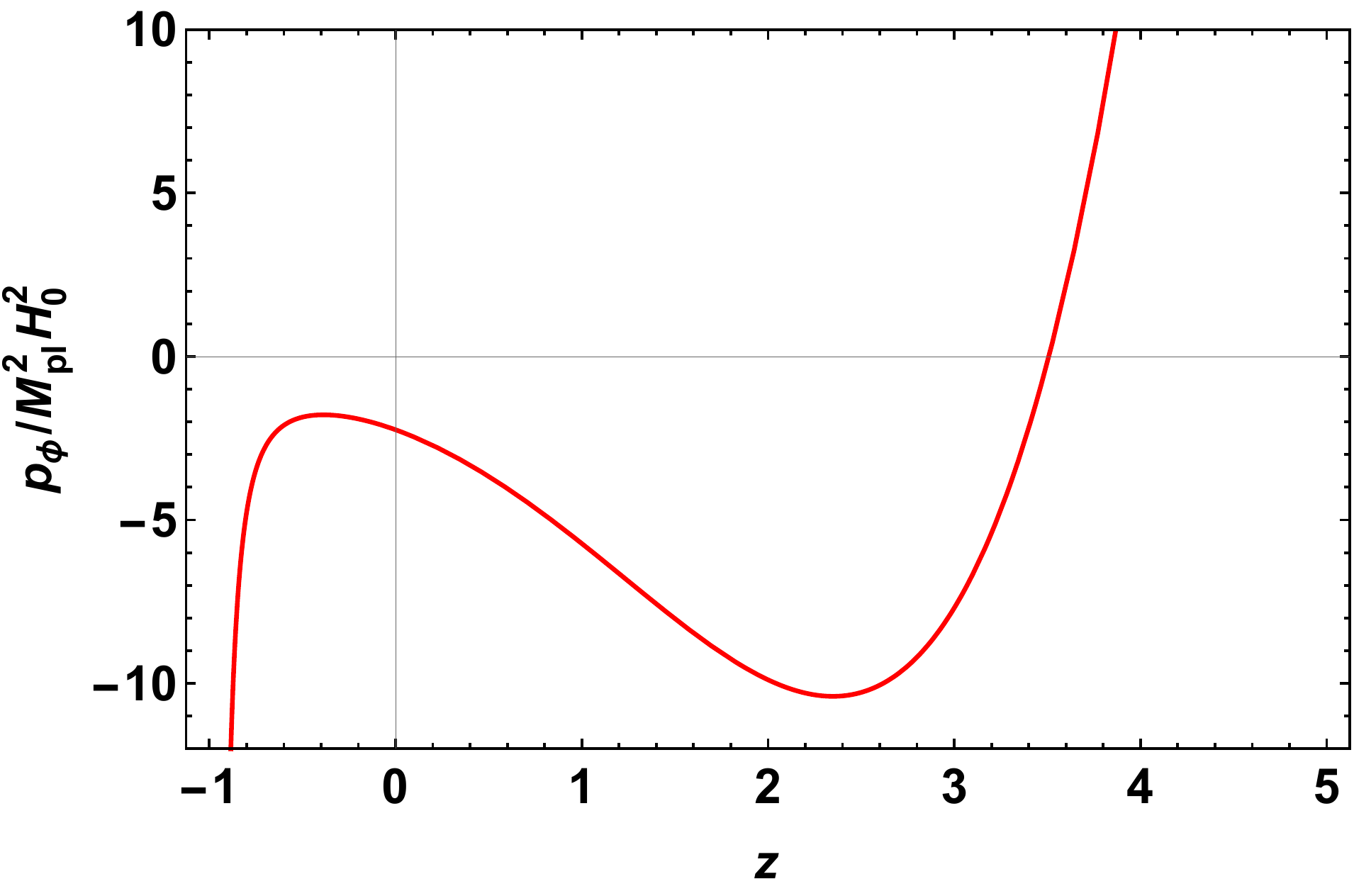}
\caption{Profile of dark energy pressure for Model 1.}\label{fig:pressurem1}
   \end{minipage}
\end{figure}

Similarly, for Model 2, we obtain,

\begin{equation}
M_{Pl}^{-2}H_{0}^{-2}\rho _{\phi } = 3[ \left( 1+\zeta
^{2\gamma }\right) ^{-3}(1+z)^{-4\gamma }\left[ 1+\{\zeta (1+z)\}^{2\gamma
}\right] ^{3}  -\Omega_{M0}(1+z)^{3}]
\end{equation}

\begin{equation}
M_{Pl}^{-2}H_{0}^{-2}p_{\phi } = \left\{ -3+2\gamma
-6\gamma \left[ 1+\{\zeta (1+z)\}^{\gamma }\right] ^{-1}\right\} \left\{ \left( 1+\zeta ^{2\gamma }\right) ^{-3}(1+z)^{-4}\left[ 1+\{\zeta
(1+z)\}^{2\gamma }\right] ^{3}\right\} 
\end{equation}

Similarly, the evolution of dark energy density and pressure for Model 2 are shown in the following figures Fig:- \ref{fig:densitym2} \& Fig:- \ref{fig:pressurem2}:

\begin{figure}[!htb]
   \begin{minipage}{0.49\textwidth}
     \centering
   \includegraphics[scale=0.42]{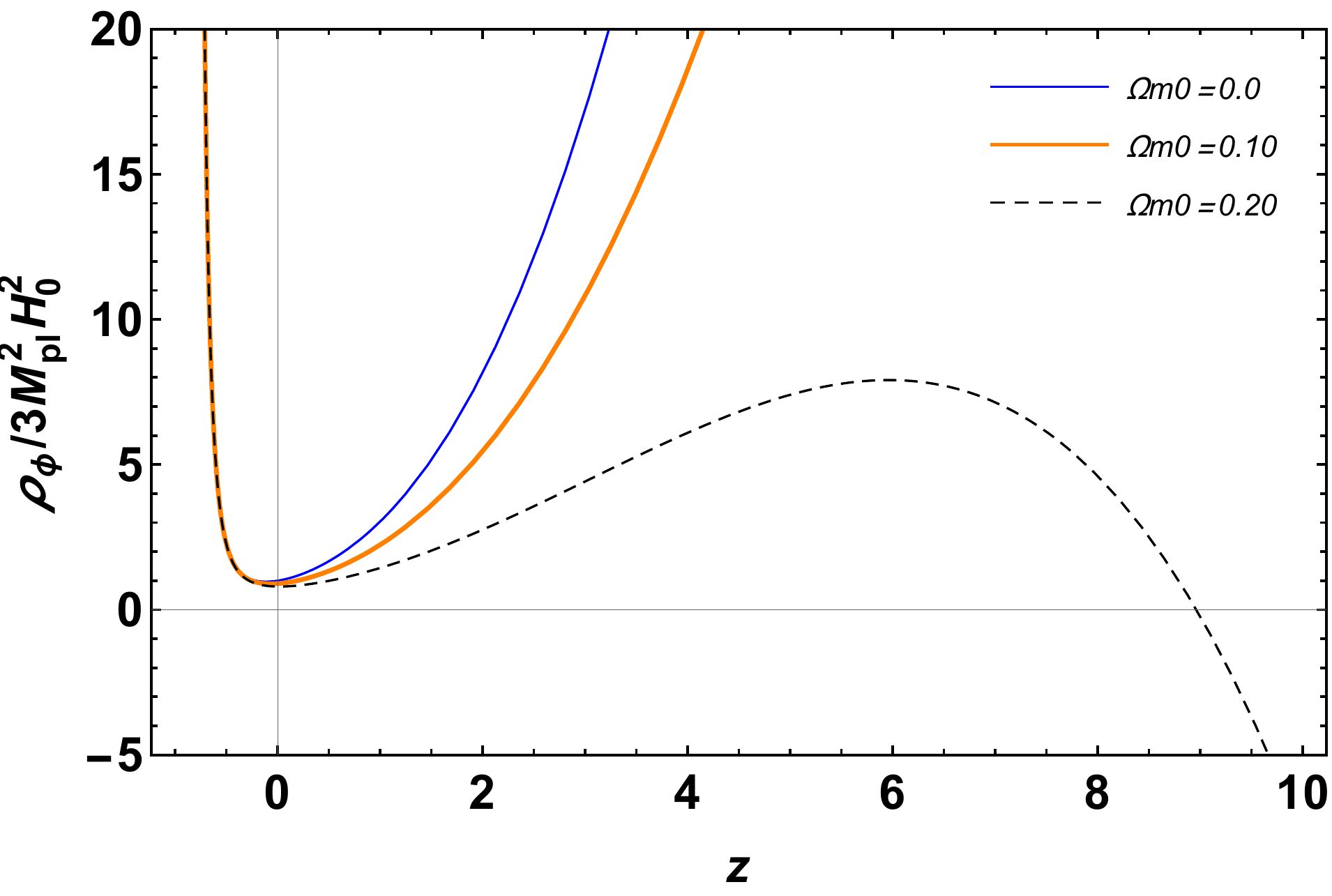}
\caption{Profile of energy density for Model 2.}\label{fig:densitym2}
   \end{minipage}\hfill
   \begin{minipage}{0.49\textwidth}
     \centering
    \includegraphics[scale=0.42]{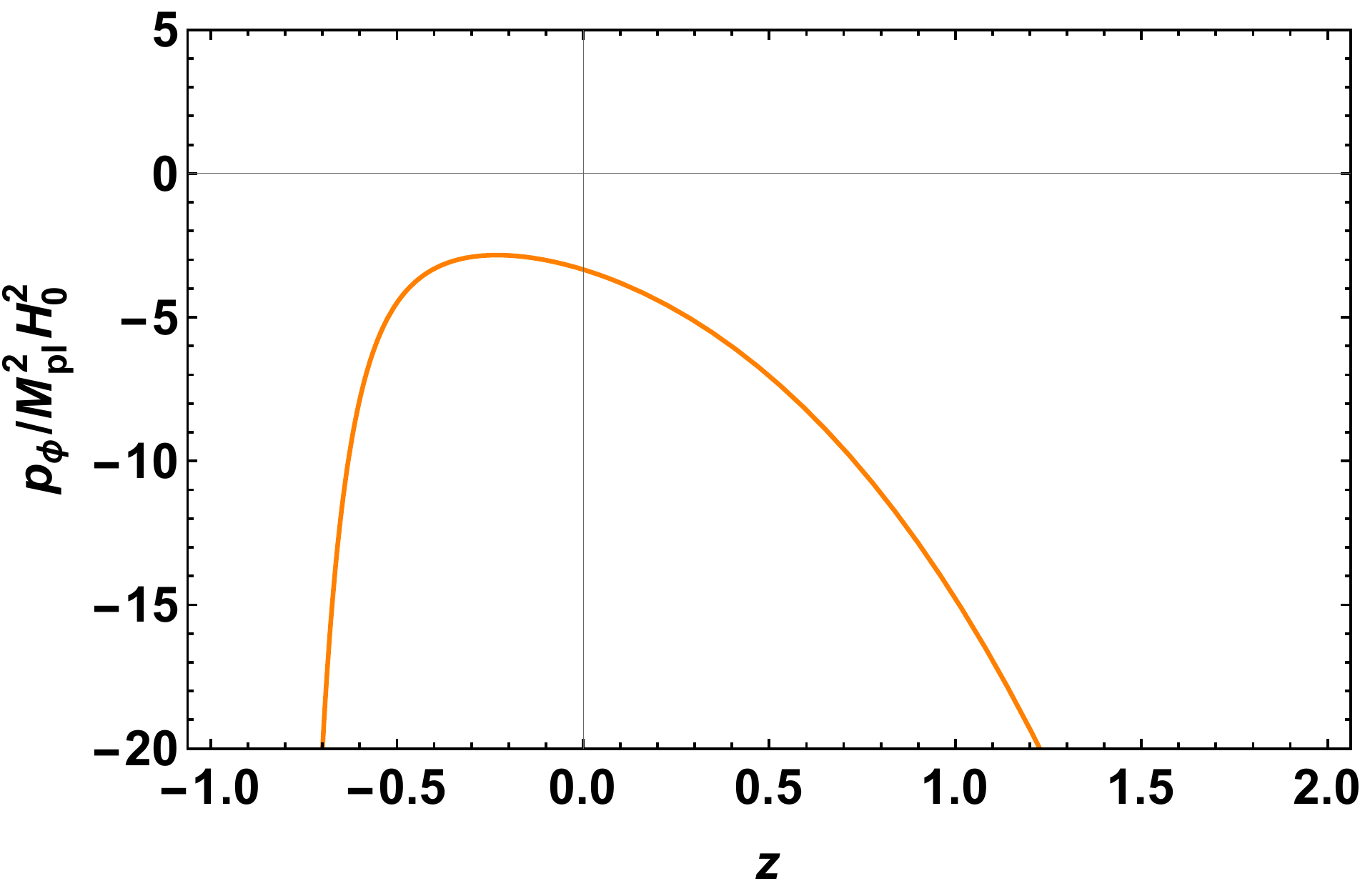}
\caption{Profile of dark energy pressure for Model 2.}\label{fig:pressurem2}
   \end{minipage}
\end{figure}


\subsection{Equation of state (EoS) parameter}

From the above calculation, it is easy to find the expressions for the equation of state parameter $\omega _{\phi }$ for both the models and the expressions are:

\begin{widetext}
	\begin{equation}
		\omega _{\phi }=\frac{\left\{ -3+2\gamma -4\gamma \left[ 1+\{\zeta
			(1+z)\}^{\gamma }\right] ^{-1}\right\} \left\{ \left( 1+\zeta ^{\gamma
			}\right) ^{-4}(1+z)^{-2}\left[ 1+\{\zeta (1+z)\}^{\gamma }\right]
			^{4}\right\} }{3\left[ \left( 1+\zeta ^{\gamma }\right) ^{-4}(1+z)^{-2\gamma
			}\left[ 1+\{\zeta (1+z)\}^{\gamma }\right] ^{4}-\Omega _{M0}(1+z)^{3}\right] 
		}  \label{s7}
	\end{equation}
	
	and
	
	\begin{equation}
		\omega _{\phi }=\frac{\left\{ -3+2\gamma -6\gamma \left[ 1+\{\zeta
			(1+z)\}^{\gamma }\right] ^{-1}\right\} \left\{ \left( 1+\zeta ^{2\gamma
			}\right) ^{-3}(1+z)^{-4}\left[ 1+\{\zeta (1+z)\}^{2\gamma }\right]
			^{3}\right\} }{3\left[ \left( 1+\zeta ^{2\gamma }\right)
			^{-3}(1+z)^{-4\gamma }\left[ 1+\{\zeta (1+z)\}^{2\gamma }\right] ^{3}-\Omega
			_{M0}(1+z)^{3}\right] }  \label{s8}
	\end{equation}
\end{widetext}

The evolution of the equation of state parameters for both Model 1 and Model 2 are shown in the following figures Fig:- \ref{fig:eosm1} \& Fig:- \ref{fig:eosm2}:

\begin{figure}[!htb]
   \begin{minipage}{0.49\textwidth}
     \centering
   \includegraphics[scale=0.42]{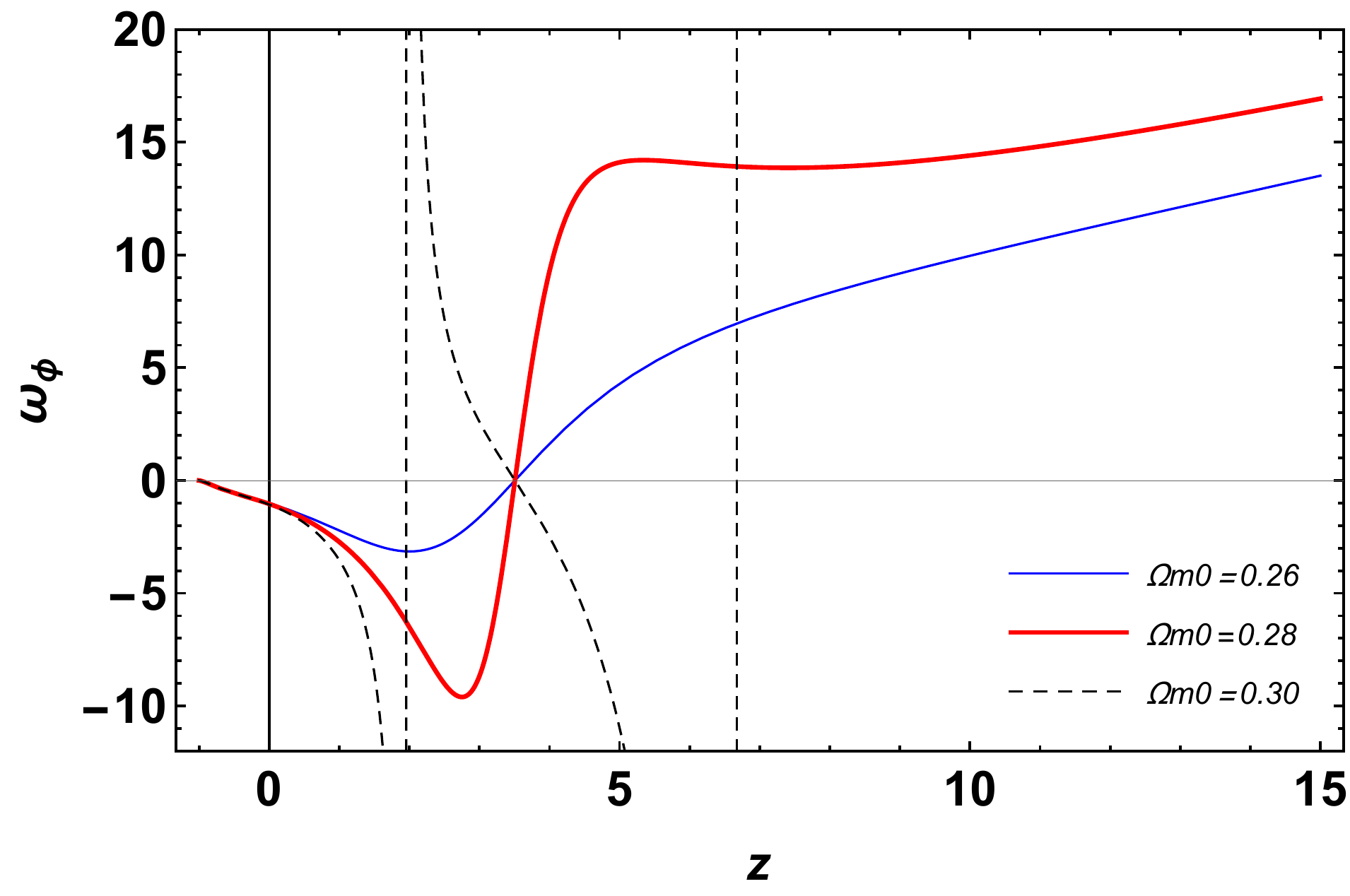}
\caption{Profile of dark energy equation of state (EoS) for Model 1}\label{fig:eosm1}
   \end{minipage}\hfill
   \begin{minipage}{0.49\textwidth}
     \centering
    \includegraphics[scale=0.42]{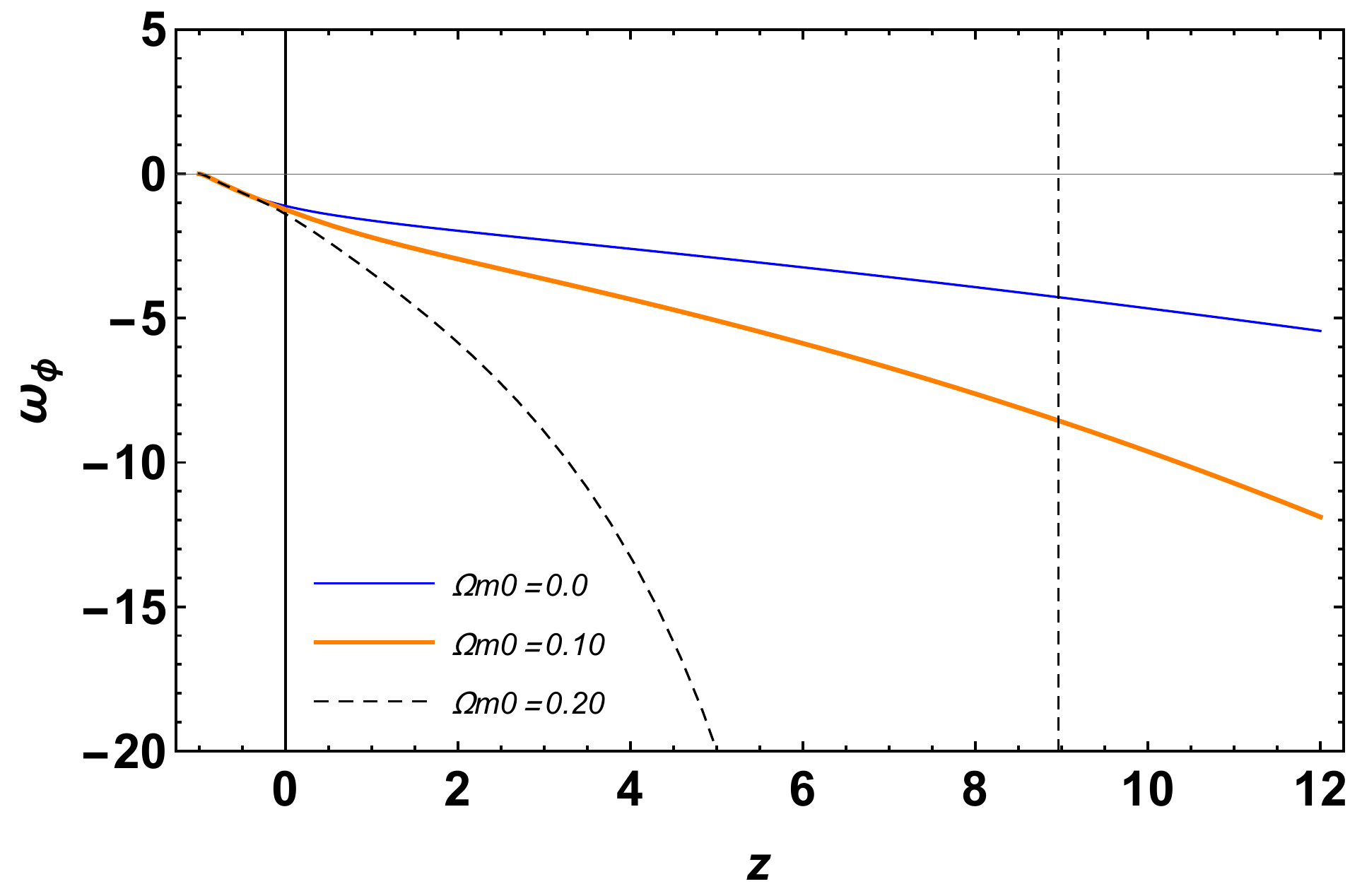}
\caption{Profile of dark energy equation of state (EoS) for Model 2}\label{fig:eosm2}
   \end{minipage}
\end{figure}

\section{Information Criteria}

In order to speak about the sustainable model analysis, one should
understand the information criteria study. (IC). The criteria for Akaike
information (AIC) \cite{84} is exclusively implemented by all ICs. Although
the AIC approximates the minimization of the Kullback-Leibler information,
it acts as an asymptotically unbiased estimator of Kullback-Leibler
information. The AIC's Gaussian estimator can be represented as \cite{85,
86, 87, 88} $\text{AIC}=-2\ln (\mathcal{L}_{max})+2\kappa +\frac{2\kappa
(\kappa +1)}{N-\kappa -1}$ in which $\mathcal{L}_{max}$ is the maximum
likelihood function, $\kappa $ is the total number of model parameters, and $%
N$ is the total number of data points employed to fit the models to the
data. Given that $N\gg 1$ for the models, the aforementioned formula changes
to the original AIC like $\text{AIC}=-2\ln (\mathcal{L}_{max})+2\kappa $.
The variations of the IC values are minimized if the collection of models is
provided. $\triangle \text{AIC}=\text{AIC}_{Model}-\text{AIC}%
_{min}=\triangle \chi _{min}^{2}+2\triangle \kappa $ Throughout data
analysis studies, the more favorable range of $\triangle \text{AIC}$ is $%
(0,2)$. The low favorable range of $\triangle \text{AIC}$ is $(4,7)$, while $%
\triangle \text{AIC}>10$ provides less support model.\\\\

\begin{table}[]
\begin{center}
\begin{tabular}{|c|c|c|c|c|}
\hline
Model & $\chi_{min}^{2}$ & $\chi_{red}^{2}$ & $AIC$ & $\Delta AIC$ \\ \hline
$\Lambda$CDM Model & 1073.67 & 0.981 & 1077.67 & 0 \\ \hline
Model 1 & 1072.89 & 0.961 & 1078.89 & 1.22 \\ \hline
Model 2 & 1074.69 & 0.963 & 1078.69 & 1.02 \\ \hline
\end{tabular}%
\end{center}
\caption{Summary of the ${{\protect\chi}^2_{min}}$, ${{\protect\chi}^2_{red}}
$, $AIC$ and $\Delta AIC$.} 
\label{table3}
\end{table}

\section{Results and Discussion}

\paragraph{deceleration parameter}
The comparison of the redshift dependence of the deceleration parameter for Model 1 and Model 2 with respect to the $\Lambda$CDM model can be understood as follows. In the redshift range $z \in (0,10)$, the evolution of the deceleration parameter appears to be similar among all three models. However, noticeable discrepancies between the models and the $\Lambda$CDM model become evident for redshifts greater than $z=1.5$. The numerical values of the transition redshift, denoted as $z_{tr}$, which marks the transition from a decelerating to an accelerating phase, are relatively close for all models, including the $\Lambda$CDM model. However, it is important to note that while the $\Lambda$CDM model ends in a de Sitter phase with a deceleration parameter of $q=-1$, both Model 1 and Model 2 exhibit super-accelerated evolution with deceleration parameters at $z=-1$ approximately equal to $-2.6542$ and $-3.85658$, respectively. These observations carry significant implications. The fact that Model 1 and 2 display super-acceleration suggests the presence of exotic or modified gravitational effects at high redshifts, beyond what is accounted for by the $\Lambda$CDM model. This behavior could be indicative of the influence of alternative theories of gravity or additional matter components with peculiar properties. The deviation of the deceleration parameter from the expected value of $-1$ in the $\Lambda$CDM model highlights the need for exploring and considering alternative cosmological models in order to fully understand the dynamics and evolution of our Universe.\\\\

\paragraph{jerk Parameter}
The behavior of the jerk parameter, denoted as $j(z)$, for Model 1 and Model 2 with respect to the standard $\Lambda$CDM paradigm can be understood as follows. It is observed that at high redshifts, the predictions of both models deviate significantly from the $\Lambda$CDM model, exhibiting opposite trends. At lower redshifts, noticeable discrepancies between the models and the $\Lambda$CDM model persist. Notably, at $z=1$, Model 1 predicts a jerk parameter value that is 1.4 times higher than the value predicted by the $\Lambda$CDM model. On the other hand, Model 2 predicts a jerk parameter value that is 13 times higher than the $\Lambda$CDM prediction. These significant deviations at lower redshifts highlight the potential for distinguishing between the different models through observational measurements of the present-day value of the jerk parameter, denoted as $j_0$. The deviations in the jerk parameter emphasize the necessity of exploring alternative models to better understand the dynamics and evolution of the Universe. The jerk parameter provides valuable insights into the nature of cosmic acceleration and the underlying physical mechanisms driving it. Therefore, precise measurements of $j_0$ can serve as a powerful tool in discriminating between different cosmological scenarios and shedding light on the fundamental properties of the Universe.\\\\

\paragraph{snap Parameter}
The behavior of the snap parameter, denoted as $s(z)$, for Model 1 and Model 2 compared to the $\Lambda$CDM model  can be understood as follows. These model's demonstrate notable discrepancies between the models and the $\Lambda$CDM model, particularly at high redshifts in the range $z\in (2,10)$. For Model 1, at lower redshifts ranging from $z\in (0,0.2)$, the discrepancies with respect to the $\Lambda$CDM model diminish. This indicates that Model 1 aligns more closely with the $\Lambda$CDM model at lower redshifts, suggesting a better agreement in terms of the snap parameter. However, at high redshifts, significant deviations emerge, indicating a systematic disparity between Model 1 and the $\Lambda$CDM model. On the other hand, Model 2 exhibits a distinct behavior for the snap parameter. In the range $z\in (0,0.2)$, Model 2 shows an increase as a function of redshift $z$, implying a slower rate of change compared to the $\Lambda$CDM model. However, at $s(0)$, the snap parameter value predicted by Model 2 is 63 times higher than the corresponding value predicted by the $\Lambda$CDM model. This substantial difference further emphasizes the contrasting nature of Model 2 compared to the standard cosmological framework. The systematic discrepancies observed at high redshifts in the snap parameter highlight the need for alternative cosmological models that can better explain the observed cosmic phenomena. By accurately measuring the snap parameter and comparing it with the predictions from different models, including Model 1 and 2, one can assess the viability and compatibility of these models with observational data. If the measured snap parameter values align more closely with the predictions of either Model 1 or Model 2 rather than the $\Lambda$CDM model, it would provide evidence for the presence of novel physical processes or the need for alternative theories of gravity.\\\\

\paragraph{Statefinder diagnostic}
The evolution of the statefinder pair $(r, s)$ for Model 1 can be understood as follows. The values in the range $r>1$ and $s<0$ correspond to a Chaplygin gas-type dark energy model. As the evolution progresses, the statefinder trajectory transitions to the quintessence region before eventually returning to the Chaplygin gas region at late times. During this evolution, the model bypasses the intermediate fixed point $(0, 1)$ associated with the $\Lambda$CDM model. $(r, q)$ profile provides additional information on Model 1 by focusing on its deviation from the de Sitter point $(-1, 1)$. The de Sitter point represents a cosmological solution characterized by constant acceleration and a constant equation of state parameters. The deviation of Model 1 from this de Sitter point signifies that the dynamics of the model exhibit deviations from a purely exponential expansion. The statefinder diagnostic allows for a more detailed examination of the dynamical properties of Model 1. The transition from a Chaplygin gas-type behavior to a quintessence and back to the Chaplygin gas region suggests a non-trivial evolution of the dark energy component throughout cosmic history. The model's deviation from the de Sitter point indicates the presence of additional dynamics and deviations from a pure exponential expansion. The evolution of the statefinder pair $(r, s)$ for Model 2 can be understood as follows. It exhibits similar behavior to Model 1, transitioning from the quintessence region in the past to the Chaplygin gas region, passing through the intermediate $\Lambda$CDM fixed point $\{0,1\}$. This behavior is further supported by $(r, q)$ profile, which also deviates from the de Sitter point $(-1,1)$, providing additional evidence for the dynamics of Model 2. These statefinder diagnostics demonstrate the distinct evolutionary characteristics of Model 2 and its deviation from the standard $\Lambda$CDM cosmology.\\\\

\paragraph{Om diagnostic} Variation of the $Om(z)$ parameter with redshift $z$ for Model 1 and 2 can be understood as follows. In both cases, the $Om(z)$ values are smaller than $\Omega _{m0}$ for $z>0$, indicating that the models reside in the quintessence domain. As the redshift decreases, the $Om(z)$ value experiences a significant decrease and becomes negative, indicating that both models enter the phantom region. This behavior highlights the transition from quintessence to phantom behavior in the evolution of Model 1 and 2, providing insights into the cosmological dynamics and characteristics of these models.\\\\

\paragraph{Energy Density \& Pressure of DE}
The cosmic evolution of dark energy density, denoted by $\rho_\phi$, and dark energy pressure for Model 1 and Model 2 provide valuable insights into the behavior and properties of quintessence dark energy in a cosmological context. The blue line and red line represent the evolution of the dark energy density for Model 1. These lines consistently maintain positive values throughout cosmic evolution. On the other hand, the dotted lines in the figure indicate some instances of negative values for the dark energy density, suggesting that the first two choices of the density parameter, as indicated by the solid lines, are more suitable. This behavior implies that Model 1, with a linearly varying deceleration parameter, can effectively describe the behavior of quintessence dark energy. Similarly , the blue line and orange line represent the evolution of the dark energy density for Model 2. These lines also exhibit positive values throughout cosmic evolution. However, unlike Model 1, Model 2 does not show instances of negative density values. While this may seem advantageous, it is important to note that the behavior of dark energy density alone does not determine the superiority of a model. Other cosmological parameters and observational constraints should be taken into account. Examining the dark energy pressure, further highlights the distinctions between Model 1 and Model 2. In both cases, Model 2 exhibits all-time negative values of dark energy pressure, indicating a consistent behavior of quintessence dark energy with negative pressure. On the other hand, Model 1 shows an early positive value of dark energy pressure, transitioning to negative values in the later stages of cosmic evolution. This behavior aligns with the expected characteristics of quintessence models, where the dark energy component initially contributes to positive pressure and subsequently drives accelerated expansion with negative pressure. Model 1, with a linearly varying deceleration parameter, provides a better description of quintessence dark energy compared to Model 2, which features a quadratic varying deceleration parameter. The positive values of dark energy density and the transition from positive to negative dark energy pressure in Model 1 align with the expected behavior of quintessence models and are indicative of a more consistent and viable cosmological scenario. These findings contribute to our understanding of the nature of dark energy and its role in cosmic evolution, offering insights for further investigations and refinements in the field of cosmology.\\\\

\paragraph{Equation of state (EoS) parameter} The behavior of dark energy density as quintessence in cosmological models carries important implications for the present-day matter density parameter, $\Omega_{m0}$. In Model 1, which exhibits a smooth evolution of the equation of state parameter, the plots suggest that $\Omega_{m0}$ should be less than 0.28 to be consistent with quintessence-like dark energy. This implies that the contribution of matter to the total energy density is relatively low compared to dark energy. In Model 2, the plots indicate that positive values of $\Omega_{m0}$ may be incompatible with quintessential dark energy. This suggests that matter alone cannot explain the observed behavior of dark energy in this model. These implications highlight the need to consider $\Omega_{m0}$ alongside dark energy behavior to ensure compatibility with observations. The values of $\Omega_{m0}$ determine the contributions of matter and dark energy, shaping the universe's evolution and expansion. Refining our understanding of $\Omega_{m0}$ and its constraints in different models enhances our knowledge of the underlying physics behind the universe's accelerated expansion and the nature of dark energy. Further investigations and observations are required to determine the precise value of $\Omega_{m0}$ and deepen our understanding of the interplay between matter and dark energy.
\\\\

\paragraph{Information Criteria}
In our analysis, we compared Model 1 and Model 2 with the $\Lambda$CDM model, evaluating their relative support using the $\Delta$AIC values. Table \ref{table3} shows that Model 1 has a $\Delta$AIC value of 1.22, while Model 2 has a $\Delta$AIC value of 1.02. These positive $\Delta$AIC values indicate that both Model 1 and Model 2 have slightly weaker support compared to the $\Lambda$CDM model. A $\Delta$AIC value within the range of (0, 2) is considered most favorable, suggesting moderate evidence in favor of a model. In our case, the $\Delta$AIC values for Model 1 and Model 2 fall within this range, indicating some degree of support for these models. Comparing the $\Lambda$CDM model to Model 1, we find that the $\Delta$AIC value for Model 1 is higher, indicating that the $\Lambda$CDM model performs slightly better in terms of goodness of fit and model selection. Similarly, when comparing the $\Lambda$CDM model to Model 2, the $\Delta$AIC value for Model 2 is higher, indicating a better fit for the $\Lambda$CDM model. Our analysis indicates that the $\Lambda$CDM model performs slightly better than Model 1 and Model 2 based on the $\Delta$AIC values.

\section{Conclusion}

We conducted a comprehensive and robust investigation of the two cosmological dark energy models in this article, contrasting them to
cosmological observations for $57$ Hubble uncorrelated measurements, the
Pantheon dataset spanning $1048$ measurements, $162$ Gamma Ray Bursts (GRBs)
measurements, $24$ measurements of compact radio quasars, and $17$
uncorrelated BAO measurements. To find the optimal values for the model
parameters. We employed the MCMC approach, which allows us to deduce the best fit for the model parameters. Using the best-fit values, the data fittings produce extremely good results for both the Hubble and the Pantheon datasets. Furthermore, there is a particularly significant relation between both the dark energy model and the $\Lambda $CDM model. The analysis of the two models has been also conducted with a more statistical sense besides studying the Akaike Information Criterion, which further demonstrates that
both the dark energy models and the standard $\Lambda $CDM model are closed
enough. We carried out a comprehensive comparison of the cosmographic
parameters to obtain a more accurate estimate of the attributes and flaws of
both cosmological models. The behavior of the deceleration parameter
appears to agree well with that of standard $\Lambda $CDM, substantial
differences often will have seemed between models when the jerk and snap
parameters are considered. As a result, the cosmographic technique can
provide the ability to distinguish different cosmological models. It also
nearly perfectly reproduces the $\Lambda $CDM model predictions, although
there are still considerable deviations for high redshifts and the values of
several cosmographic parameters. In terms of interpretations of empirical
observations, both the cosmological model and the conventional $\Lambda $CDM
model might become acceptable mathematical alternatives. It may also yield
novel insights as well as a decent understanding of the complicated
relationship between mathematical concepts structures and physical reality.\\\\

\textbf{Acknowledgement: }Author SKJP thanks IUCAA, Pune for hospitality and other facilities under its IUCAA associateship program, where a large part of work has been done.\\\\

\textbf{Author contributions:} The data analysis, several graphical
representations in the article, and the first draft of the manuscript were all completed by the authors AB and HC. The writing and a few computations were supplied by author AM. Project administrator and contributor SKJP wrote the document, completed the project, and contributed some computations. The  paper has been read and approved by all authors.\\\\ 

\textbf{Funding:} There is no fund available for the publication of this research article.\\\\

\textbf{Data Availability Statement:} This manuscript has used publicly
available data for the work.\\\\

\textbf{Conflict of interest:} The authors have no relevant financial or non-financial interests to disclose.\\\\

\textbf{Ethical statements:} The submitted work is original and has not been published anywhere else.

\end{document}